\documentclass[useAMS,usenatbib]{mn2e}
\usepackage{latexsym}
\usepackage{amsmath}
\usepackage{amssymb}
\usepackage{fnpos}
\usepackage{scrextend}
\usepackage{scalerel}
\usepackage{hyperref}
\usepackage{tabularx}
\usepackage{xspace}
\usepackage{enumerate}
\usepackage{graphicx}
\usepackage{caption}
\usepackage{gensymb}
\usepackage{epstopdf}
\usepackage{longtable}
\usepackage{lscape}
\usepackage{subcaption}
\usepackage{pdfpages}

\newcommand{\eal}[2]{\ifmmode{\mathrm{#1\,#2}}\else{#1\textsc{$\,$\lowercase{#2}}}\fi\xspace}
\newcommand{\feal}[2]{\ifmmode{\mathrm{#1\,#2}}\else{[#1\textsc{$\,$\lowercase{#2}}]}\fi\xspace}
\newcommand{\hfeal}[2]{\ifmmode{\mathrm{#1\,#2}}\else{#1\textsc{$\,$\lowercase{#2}}]}\fi\xspace}

\title[nova SMCN 2016-10a]{Multiwavelength observations of nova SMCN 2016-10a ---  one of the brightest novae ever observed}

\author[Aydi et al.]{E. Aydi$^{1,2}$\thanks{E-mail: eaydi@saao.ac.za}, K. L. Page$^{3}$, N. P. M. Kuin$^{4}$, M. J. Darnley$^{5}$, F. M. Walter$^{6}$, P. Mr{\'o}z$^{7}$, 
\newauthor  D. Buckley$^{1}$, S. Mohamed$^{1,2}$, P. Whitelock$^{1,2}$, P. Woudt$^{2}$, S. C. Williams$^{8,5}$, M. Orio$^{9,10}$, 
\newauthor R. E. Williams$^{11}$,  A. P. Beardmore$^{3}$, J. P. Osborne$^{3}$, A. Kniazev$^{1,12,13}$  \newauthor V. A. R. M. Ribeiro$^{14,15,16}$, A. Udalski$^{7}$, J. Strader$^{17}$ and L. Chomiuk$^{17}$\\\\
$^{1}$South African Astronomical Observatory, P.O. Box 9, 7935 Observatory, South Africa\\
$^{2}$Astronomy Department, University of Cape Town, 7701 Rondebosch, South Africa\\
$^{3}$X-ray and Observational Astronomy Group, Department of Physics \& Astronomy, University of Leicester, LE1 7RH, UK\\
$^{4}$University College London, Mullard Space Science Laboratory, Holmbury St. Mary, Dorking RH5 6NT, UK\\
$^{5}$Astrophysics Research Institute, Liverpool John Moores University, IC2 Liverpool Science Park, Liverpool, L3 5RF, UK\\
$^{6}$Department of Physics and Astronomy, Stony Brook University, Stony Brook NY 11794-3800, USA\\
$^{7}$Warsaw University Observatory, Al. Ujazdowskie 4, 00-478 Warsaw, Poland\\
$^{8}$Physics Department, Lancaster University, Lancaster, LA1 4YB, UK \\
$^{9}$INAF--Osservatorio di Padova, vicolo dell’ Osservatorio 5, I-35122 Padova, Italy\\
$^{10}$Department of Astronomy, University of Wisconsin, 475 N. Charter Str., Madison, WI 53704, USA\\
$^{11}$Space Telescope Science Institute, 3700 San Martin Drive, Baltimore, MD  21218  USA \\
$^{12}$Southern African Large Telescope Foundation, PO Box 9, 7935 Observatory, Cape Town, South Africa\\
$^{13}$Special Astrophysical Observatory of RAS, Nizhnij Arkhyz, Karachai-Circassia 369167, Russia\\
$^{14}$CIDMA, Departamento de F\'isica, Universidade de Aveiro, Campus de Santiago, 3810-193 Aveiro, Portugal\\
$^{15}$Instituto de Telecomunica\c{c}\~{o}es, Campus de Santiago, 3810-193 Aveiro, Portugal\\
$^{16}$Department of Physics and Astronomy, Botswana International University of Science \& Technology, Private Bag 16, Palapye, Botswana\\
$^{17}$Center for Data Intensive and Time Domain Astronomy, Department of Physics and Astronomy, Michigan State University, East Lansing, \\
MI 48824, USA\\
}

\begin{document}

\date{Accepted: 2017 October 10 . Received ***; in original form: 2017 July 19}
\pagerange{\pageref{firstpage}--\pageref{lastpage}} \pubyear{2015}
\maketitle

\label{firstpage}
\begin{abstract}
We report on multiwavelength observations of nova SMCN 2016-10a. The present observational set is one of the most comprehensive for any nova in the Small Magellanic Cloud, including: low, medium, and high resolution optical spectroscopy and spectropolarimetry from SALT, FLOYDS, and SOAR; long-term OGLE $V$- and $I$- bands photometry dating back to six years before eruption; SMARTS optical and near-IR photometry from $\sim$ 11 days until over 280 days post-eruption; {\em Swift} satellite X-ray and ultraviolet observations from $\sim$ 6 days until 319 days post-eruption. The progenitor system contains a bright disk and a main sequence or a sub-giant secondary. The nova is very fast with $t_2 \simeq$ 4.0\,$\pm$\,1.0\,d and $t_3 \simeq$ 7.8\,$\pm$\,2.0\,d in the $V$-band. If the nova is in the SMC, at a distance of $\sim$\,61\,$\pm$\,10\,kpc, we derive $M_{V,\mathrm{max}} \simeq - 10.5$\,$\pm$\,0.5, making it the brightest nova ever discovered in the SMC and one of the brightest on record. At day 5 post-eruption the spectral lines show a He/N spectroscopic class and a FWHM of $\sim$\,3500\,km\,s$^{-1}$ indicating moderately high ejection velocities. The nova entered the nebular phase\,$\sim$\,20 days post-eruption, predicting the imminent super-soft source turn-on in the X-rays, which started $\sim$ 28 days post-eruption. The super-soft source properties indicate a white dwarf mass between 1.2\,M$_{\odot}$ and 1.3\,M$_{\odot}$ in good agreement with the optical conclusions.
\end{abstract}


\section{Introduction}
\label{Intro}
A classical nova (CN) is a result of an eruption on the surface of a white dwarf (WD) in an interacting binary, namely ``cataclysmic variable''. Classical novae (CNe) consist of an accreting WD and a companion that typically fills its Roche lobe \citep{Warner_1995}. The accreted matter builds up on the surface of the WD and when the pressure and density reach critical levels, a thermonuclear runaway occurs \citep{Payne-Gaposchkin_1964}. Within a few days, the eruption increases the visual brightness of the star between 8 and 15 mag. After the peak, the brightness of the star decreases gradually over a timescale of a few weeks up to months. Typically, the ejecta from the eruption can reach a velocity of $\sim$ 1000\,$\mathrm{km\,s^{-1}}$ and contains a mass between $10^{-6}\,\mathrm{M_{\odot}}$ and $10^{-4}\,\mathrm{M_{\odot}}$ \citep{Payne-Gaposchkin_1957,Gallaher_etal_1978}. Although more recently faster, lower ejecta mass systems have been found \citep{Kasliwal_etal_2011,Shara_etal_2017_II}.
\\

Immediately following the thermonuclear runaway, the surface of the WD is obscured by material thrown out during the eruption. Over time these ejecta expand, becoming optically thin to X-rays, allowing the atmosphere of the nuclear burning region on the WD to become visible. This emission typically peaks in the soft X-ray band, when the nova then enters what is known as the super-soft source (SSS) state \citep{Krautter_etal_1996} (for a review on X-ray observations of CNe see e.g. \citealt{Osborne_2015} and references therein).

Novae are also observed in ultraviolet (UV) light. In some cases, such as V2491 Cyg \citep{Page_etal_2010} and V745 Sco \citep{Page_etal_2015}, the UV light-curve is completely uncorrelated with the X-ray emission, indicating the emission regions are unrelated. HV Cet \citep{Beardmore_etal_2012} is an example where the X-ray and UV emission varies in phase throughout the orbit; in this case, we may be observing a high-inclination (edge-on disk) system, where the X-ray- and UV-emitting region is periodically occulted, perhaps by the lip of an accretion disc. The final situation is one in which the X-rays and UV are anti-correlated -- for example, V458 Vul \citep{Ness_etal_2009,Schwarz_etal_2011}. This could be explained by temperature variations \citep{Schwarz_etal_2011}, with the spectral energy distribution shifting between the extreme-UV (EUV) and X-ray bands as the photosphere expands and contracts (see also \citealt{Schwarz_etal_2015}).\\

\begin{figure*}
\begin{center}
  \includegraphics[width=\textwidth]{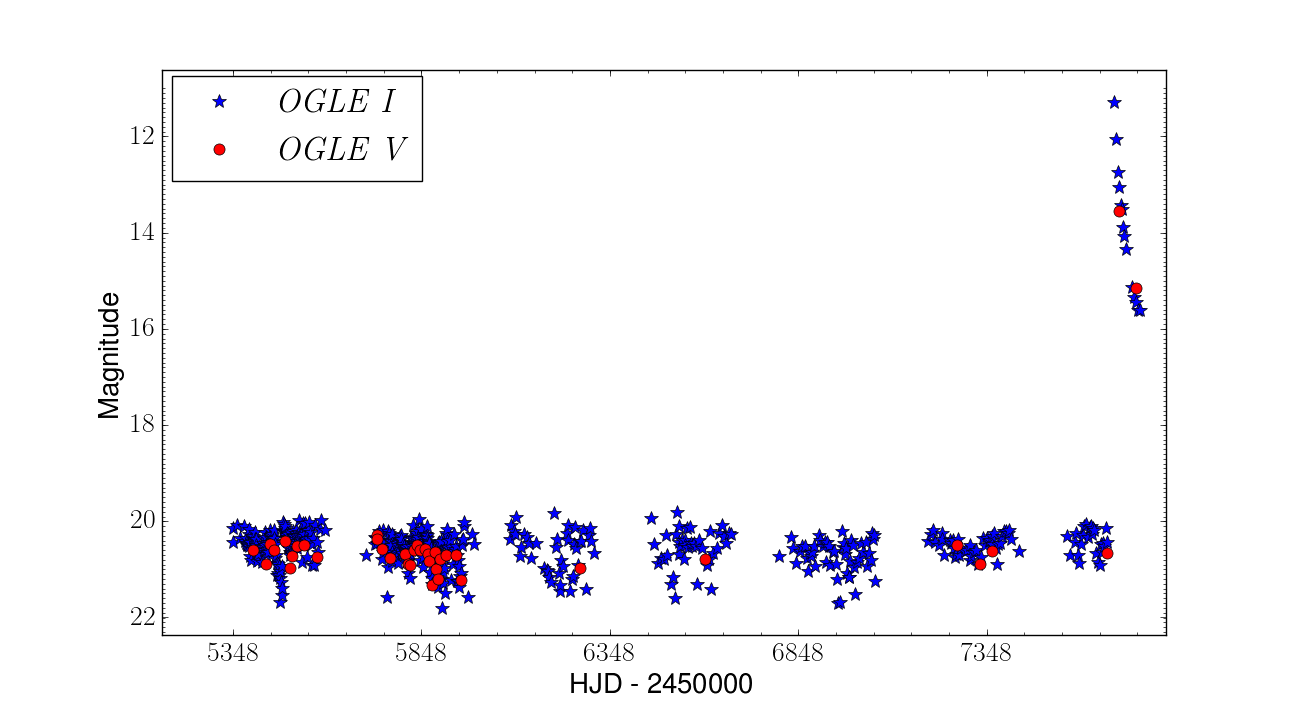}
\caption{The photometric $VI$ data, from the OGLE-IV Survey \citep{Udalski_etal_2015}, as a function of Heliocentric Julian Date (HJD), colour and symbol coded as indicated in the legend. The standard error on individual observations in quiescence is less than 0.2 mag, so much of the apparent variability is real.}
\label{Fig:OGLE_LC}
\end{center}
\end{figure*}

Novae are classified based on their optical light-curves and spectra. This classification is determined by the duration in days for the brightness to decrease by 2 or 3 mag from the intensity peak, known as $t_2$ and $t_3$, respectively \citep{Payne-Gaposchkin_1964,Shafter_1997}. \citet{Strope_etal_2010} classified a sample of 93 nova light-curves based on visual estimates and $V$ band photometry from the AAVSO Photometric All-Sky Survey (APASS \footnote{https://www.aavso.org/apass} - \citealt{Henden_etal_2009}). 

From a spectroscopic point of view novae are classified according to the early decline spectra, before they enter the nebular phase \citep{McLaughlin_1944,Payne-Gaposchkin_1957,Williams_2012}. Based on the post-eruption spectral features, \citet{Williams_1992} divided novae into two spectroscopic classes ``\eal{Fe}{II}'' and ``He/N''. Aside from the Balmer lines that dominate the spectrum in all classes, the first class exhibits narrow \eal{Fe}{II} emission lines with P Cygni profiles and the second one exhibits broad He and N emission lines. Some novae show features from both classes simultaneously or a transition from one class to the other and they are classified as hybrids \citep{Williams_1992,Williams_Bob_1994,Williams_2012}. The spectra of the \eal{Fe}{II} class may be formed in the circumbinary envelope which originated from mass loss from the secondary. The intensity of \eal{Fe}{II} spectra fades slowly and lasts for weeks and sometimes months after the eruption due to the low expansion velocities. The spectra of the He/N class are thought to originate from the gas on the WD due to the prominence of He and N transitions, the rectangular line profiles, and the large line widths \citep{Williams_2012}.\\
\\

Novae in the Milky Way and M31 have been extensively studied with global nova rates of $\sim$ 34$^{+15}_{-12}$ yr$^{-1}$ and $\sim$ 65$^{+16}_{-15}$ yr$^{-1}$, respectively \citep{Darnley_etal_2006}. In contrast, novae in the Small Magellanic Cloud (SMC) occur less frequently. Since 1897 only 22 novae have been discovered in the SMC (\citealt{Pietsch_2010}\footnote{\url{https://mpe.mpg.de/~m31novae/opt/smc/index.php}}), and most of these novae have been poorly investigated with multiwavelength observations. Little effort has been made to constrain the nova rate for the SMC (see e.g. \citealt{Graham_1979,Della_Valle_2002}). Recently \citet{Mroz_etal_2016} published an atlas of classical novae in the Magellanic Clouds, offering a systematic study of novae in the SMC where they derived an observed rate of  0.5\,$\pm$\,0.2\,yr$^{-1}$ and an absolute rate of 0.9\,$\pm$\,0.4\,yr$^{-1}$.
With such a low nova rate, multiwavelength studies of nova events in the SMC are of interest for: (1) understanding better the nova event itself, (2) providing more insights into nova properties in parent galaxies that have different metallicity content, luminosity classes, Hubble types, and star formation histories from the Milky Way or M31 \citep{Mason_etal_2005,Mroz_etal_2016}, and (3) the distance modulus to novae belonging to the SMC is known more accurately than is typical for Galactic novae, so that physical properties can also be derived more accurately.\\

Tracking the evolution of the post-eruption spectral lines and relating this  to the multiwavelength light-curve evolution is an important tool to understand the physical mechanism of nova eruptions, hence the importance of extensive low and high resolution spectroscopic monitoring of novae. High resolution spectroscopy allows us to study the evolution of line profiles which provides details  about the morphology and velocities of the ejecta, and also offers us the opportunity for modelling the system \citep{Ribeiro_etal_2013_apj,Ribeiro_etal_2013,De_Gennaro_Aquino_etal_2015,Jack_etal_2017}.

One of the advantages of studying novae in the Magellanic clouds (MCs) is the known distance (albeit with a finite spread). Deriving distances to novae is a challenge. High ejecta velocities and different light-curve decline rates (e.g. \citealt{Strope_etal_2010}) make it difficult to derive an accurate estimate. Nevertheless, many methods have been suggested for this purpose. Most of these methods are based on the decline rate of the nova optical light brightness which is correlated with their absolute magnitude at maximum light and are known as ``Maximum Magnitude versus Rate of Decline'' (MMRD) relations (e.g. \citealt{McLaughlin_DB_1939,Livio_1992,Della_Valle_Livio_1995,Downes_etal_2000}). Recently, these methods have been questioned (e.g. \citealt{Kasliwal_etal_2011,Cao_etal_2012, Shara_etal_2017_II}). Alternative methods have been proposed to derive distances to novae (see e.g. \citealt{Downes_etal_2000,Shara_etal_2017_III}) and we will refer to these methods in Section~\ref{dist}.\\

\begin{figure*}
\begin{center}
  \includegraphics[width=\textwidth]{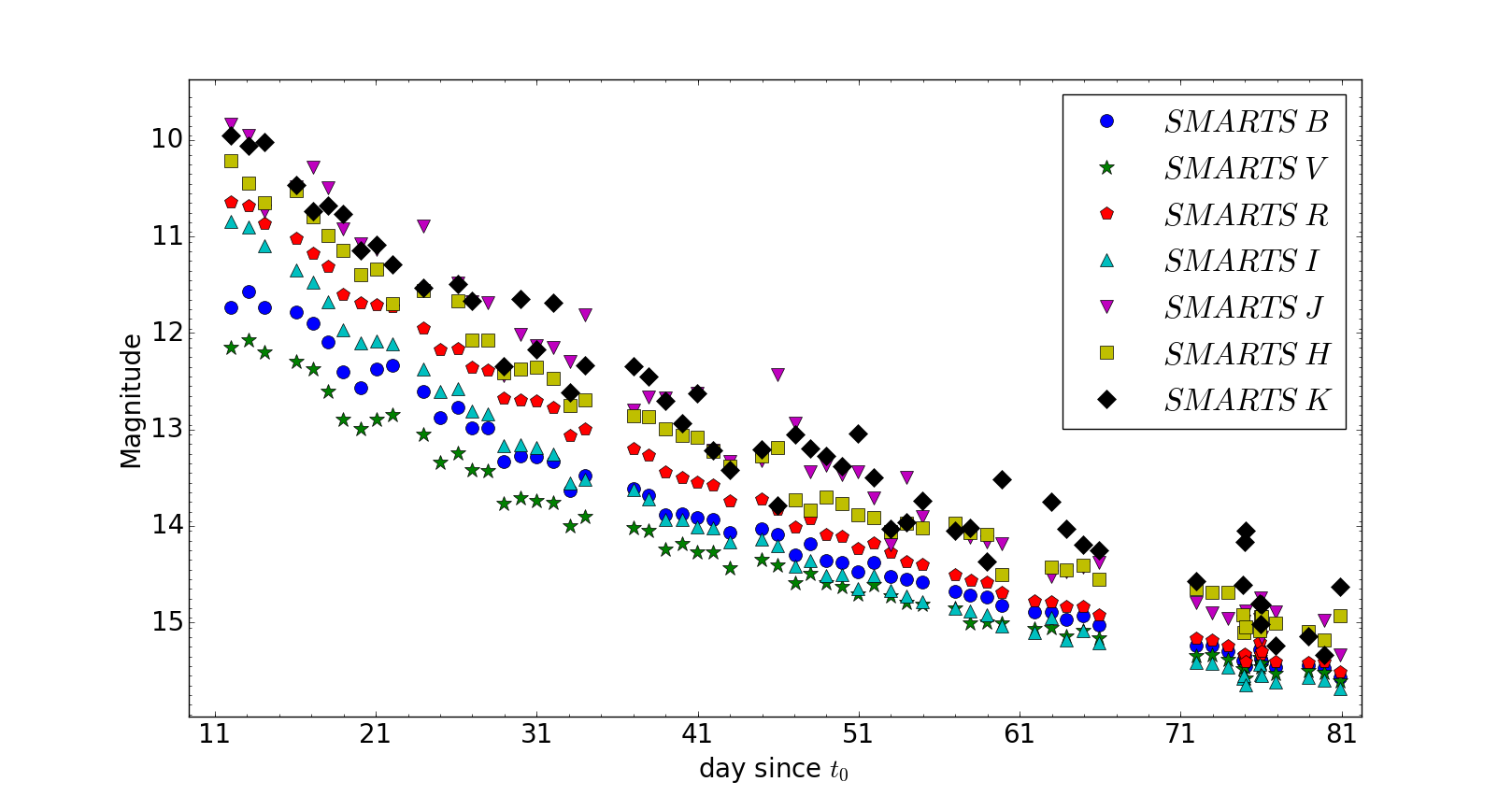}
\caption{The photometric $BVRIJHK$ data from SMARTS as a function of days since eruption, colour and symbol coded as indicated in the legend. The scatter at K is due to observational uncertainty. Data after day $\sim$ 80 suffer from moderate uncertainties (especially in the $JHK$ bands) and thus are not presented in the plot. A colour version of this plot is present in the online journal.}
\label{Fig:LC}
\end{center}
\end{figure*}

In this paper we report on multiwavelength spectroscopic and photometric observations of nova SMCN 2016-10a that was discovered by MASTER (OT J010603.18-744715.8) on  HJD 2457675.6 (2016-10-14.2 UT;\citealt{ATel_9621}). The eruption started on HJD 2457670.7 (2016-10-09.2 UT; \citealt{ATel_9684}), this date is considered as $t_0$ throughout the paper. The nova is located in the direction of the SMC at equatorial coordinates, ($\alpha$, $\delta$)$_{J2000.0}$ = (01$^{\mathrm{h}}$06$^{\mathrm{m}}$03$^{\mathrm{s}}$\!\!\!.\,27, $-$74$^{\circ}$47${\arcmin}$15${\arcsec}$\!\!\!.\,8) \citep{ATel_9622}. It might be the brightest nova ever discovered in the SMC and one of the brightest on record. Multiwavelength follow-up has been carried out to study in detail the post-eruption behaviour of the nova at optical, near-infrared (NIR), X-ray, and UV wavelengths. 

The paper is outlined as follows: in Sections~\ref{photo} and~\ref{spec} we present and analyze the optical photometric and spectroscopic observations, respectively. We present the X-ray data in Sections~\ref{xray} and the UV data in Section~\ref{UV}, followed by our discussion in Section~\ref{Disc}. We present a summary and the conclusions in Section~\ref{Conc}. The observation log is presented in Appendix~\ref{app}.


\section{Optical photometric observations and data reduction}  
\label{photo}
\subsection{OGLE observations and data reduction}
The nova has been monitored since 2010 in the $I$- and $V$-bands by the OGLE-IV Survey \citep{Udalski_etal_2015}, as part of the Magellanic System survey. These long-term observations provide information on the progenitor system. All data were reduced and calibrated following the standard OGLE pipeline \citep{Udalski_etal_2015}. 
Owing to saturation, the data poorly cover the eruption period and the early decline. The last observations before the eruption were made on HJD 2457667.6 (2016 October 6) and the first unsaturated image was obtained on HJD 2457686.6 (2016 October 25). The OGLE $V$- and $I$- band data are presented in Figure~\ref{Fig:OGLE_LC}.

\subsection{SMARTS observations and data reduction}
The Small and Moderate Aperture Research Telescope System (SMARTS) obtained optical {\em BVRI} and near-infrared (NIR) {\em JHK} photometric observations. The observations started on HJD 2457682.6 (2016 October 21). The integration times at {\em JHK} were 15\,s (three 5\,s dithered images) before 2016 December 03, and 30\,s thereafter. Optical observations are single images of 30, 25, 20, and 20\,s integrations respectively in {\em BVRI} prior to 2016 December 03, and uniformly 50\,s after 2016 December 03. The details of the data reduction are presented in \citet{Walter_etal_2012}. The {\em BVRI} and {\em JHK} data are plotted in Figure~\ref{Fig:LC}.

\begin{figure*}
\begin{center}
 \includegraphics[width=\textwidth]{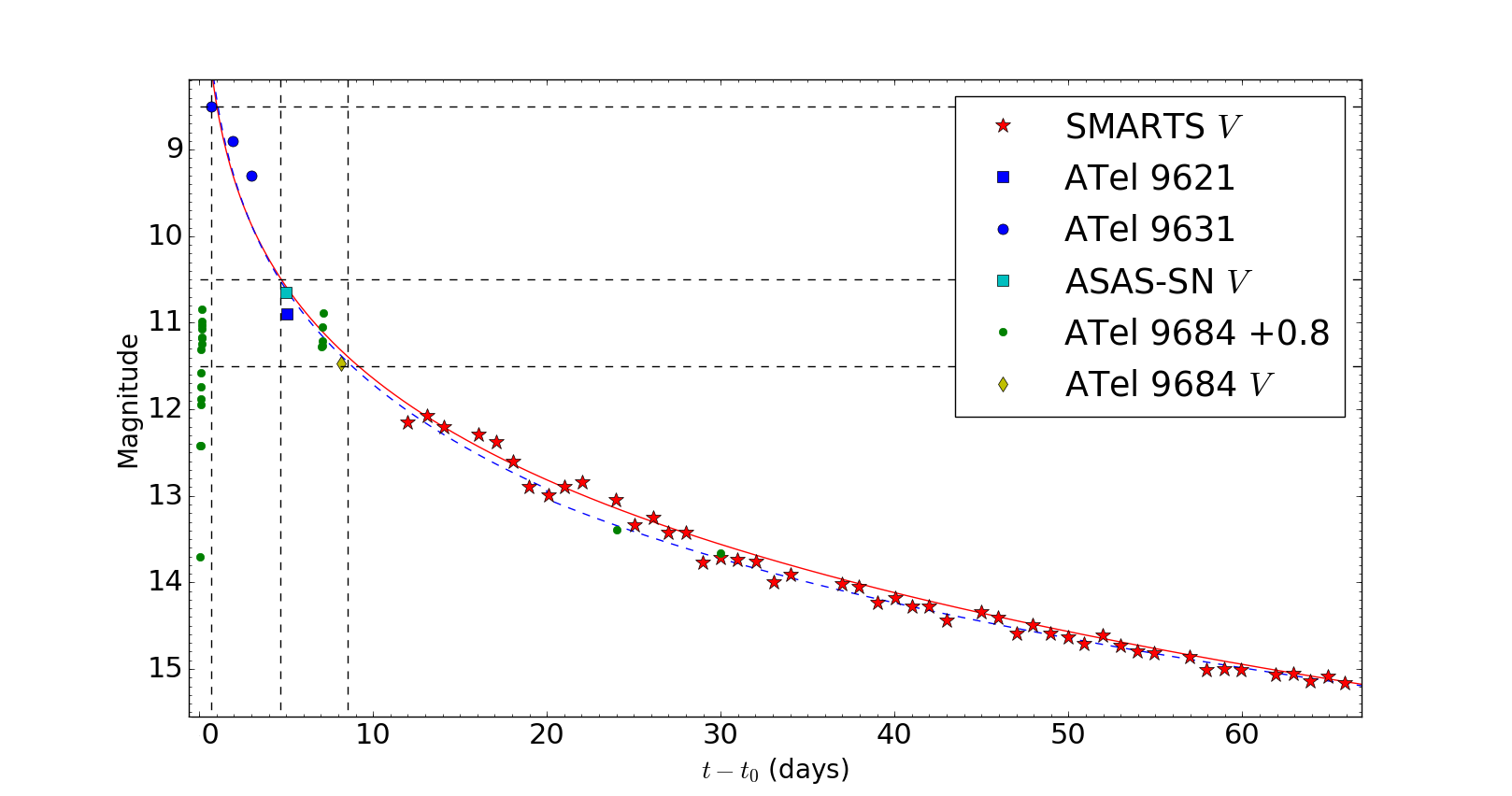}
\caption{$V$-band photometry from SMARTS and ASAS-SN along with data from \citet{ATel_9621,ATel_9631,ATel_9684}. An offset of + 0.8 is added to the data from (\citealt{ATel_9684}; see main body of text). The blue dashed line represents a broken power law fit to the data and the red solid line represents a power law fit to the data. The black horizontal dashed lines represent from top to bottom ($V_{\mathrm{max}}$), ($V_{\mathrm{max}} + 2$), and ($V_{\mathrm{max}} + 3$) respectively. The black vertical dashed lines represent from left to right $t_{\mathrm{max}}$, $t_2$, and $t_3$ respectively.}
\label{Fig:best_fist}
\end{center}
\end{figure*}

\subsection{Optical light-curve parameters}
\label{LC}
CNe light-curves are characterized by several parameters: the rise rate, the rise time to maximum light, the maximum light, the decline rate, and the decline behaviour (morphology) (e.g. \citealt{Hounsell_etal_2010,Cao_etal_2012}). The light-curves of nova SMCN 2016-10a show a smooth decline with weak oscillations. In comparison to the \citet{Strope_etal_2010} classification, the light-curves of nova SMCN 2016-10a look like an S-class event. The S-class is considered as the stereotypical nova light-curve and it is characterized by a rapid rise, a relatively sharp peak, followed by a smooth decline that starts off steep and then becomes more gradual with time \citep{Strope_etal_2010}. Note that the SMARTS optical photometry show a plateau starting about day $\sim$ 100 and ending after day $\sim$ 130, during solar conjunction (see Figure~\ref{Fig:BVRI_LC}).

On HJD 2457667.67 the nova was observed with OGLE $V$-band at a magnitude of 20.65. Pre-discovery observations by a robotic DSLR camera at Sao Jose dos Campos, Brazil \citep{ATel_9684} showed that on HJD 2457670.7 (2016-10-09.2 UT) the nova was at a magnitude of 12.9. We assume this date as $t_0$. 

The magnitude then reached 9.9, after 0.1\,days. The nova subsequently brightened to a maximum on HJD 2457671.3 at a magnitude of 8.5. After 1.2\,days the magnitude dropped to 8.9, then to 9.3 two days later, indicating the start of the decline. These measurements were obtained using the MASTER Very Wide Field (VWF) camera. According to \citet{Gorbovskoy_etal_2010}, the instrumental photometric band of the MASTER VWF camera can be described fairly well by the $V_{\mathrm{TYCHO2}}$ filter and it corresponds to $V$-band photometry (private communication with E. Gorbovskoy). We will consider these measurements as an approximation of $V$-band photometry. \citet{ATel_9621} discovered the nova at an unfiltered magnitude of 10.9 on HJD 2457675.6.

We consider HJD 2457671.3 (2016-10-9.8 UT) as the day  of maximum light and $V_{\mathrm{max}}$\,$\approx$ 8.5. This means that the magnitude rose from $>$\,13 to 8.5 in less than 1\,day which is relatively fast compared to other classical novae (see e.g. \citealt{Hounsell_etal_2010,Cao_etal_2012}).

In order to get an estimate of the rate of decline, we carried out broken power law and power law fitting to the MASTER and SMARTS magnitudes. We also make use of the All-Sky Automated Survey for Supernovae (ASAS-SN, \citealt{Shappee_etal_2014}) $V$-band measurement 1.7 h before the discovery. A broken power law is best fitting an S-class light-curve (see e.g. \citealt{Hachisu_etal_2006,Strope_etal_2010}).
The fitting resulted in $t_2\simeq$ 4.0\,$\pm$\,1.0\,d, $t_3\simeq$ 7.8$\,\pm$\,2.0\,d, and a decline rate of $\simeq 0.55 \pm 0.1$\,mag\,d$^{-1}$ over $t_2$. A power law index of $\sim$ 1.76 was derived from the early slope, in good agreement with the theoretical power law of \citet{Hachisu_etal_2006} and the empirical values of \citet{Strope_etal_2010}. We also derive $t_2$ and $t_3$ from a linear interpolation between points and we get very similar results.

We increased the uncertainty on $t_2$ and $t_3$ to take into consideration the usage of different measurements from different telescopes and detectors. It is important to recognize that several of our conclusions are dependent on these assumptions. By comparing to the empirical relation from \citet{Hachisu_etal_2006} which relates $t_3$ to $t_2$ as: $t_3 = 1.69t_2 + 0.69 \Delta t_0$ days, with the rise time $\Delta t_0 \sim$ 0.6, we obtain $t_3 = 7.1\pm1.0$, in agreement with the values estimated above.

In Figure~\ref{Fig:best_fist} we present the first 70 days of the SMARTS $V$-band photometry along with data from \citet{ATel_9621,ATel_9631,ATel_9684} and the $V$-band measurement from ASAS-SN. The DSLR cameras used to obtain the data from \citet{ATel_9684} are very red sensitive and the data require colour correction to be compared with $V$-band photometry. Because of the colour correction required, we exclude these data from the fitting and we just plot them for comparing their agreement with the fit. We adopt an offset of + 0.8 mag from \citet{Munari_etal_2017} to correct for the colour difference. As can be seen in Figure~\ref{Fig:best_fist}, the broad DSLR measurement after maximum agree with the simultaneous SMARTS $V$-band data. 

\citet{Munari_etal_2017} derived different values of $t_2$ and $t_3$. This is because of the offset they added to the MASTER VWF data \citep{ATel_9631}. They considered that the VWF is very red sensitive and they added an offset of +1 mag to correct for the colour difference, leading them to larger values of $t_2$, $t_3$, and $V_{\mathrm{max}}$. Since the MASTER VWF measurements correspond to $V$-band photometry (Gorbovskoy, private communication), such an offset is too large. 

\citet{Payne-Gaposchkin_1964} classified the light-curves of novae into 5 speed classes. Novae with $t_2$\,$<$\,10\,days and $\dot{m}_v$\,$>$\,0.20\,mag\,d$^{-1}$, which is the case for this nova, are classified as ``very fast". 
\begin{figure*}
  \includegraphics[width=\textwidth]{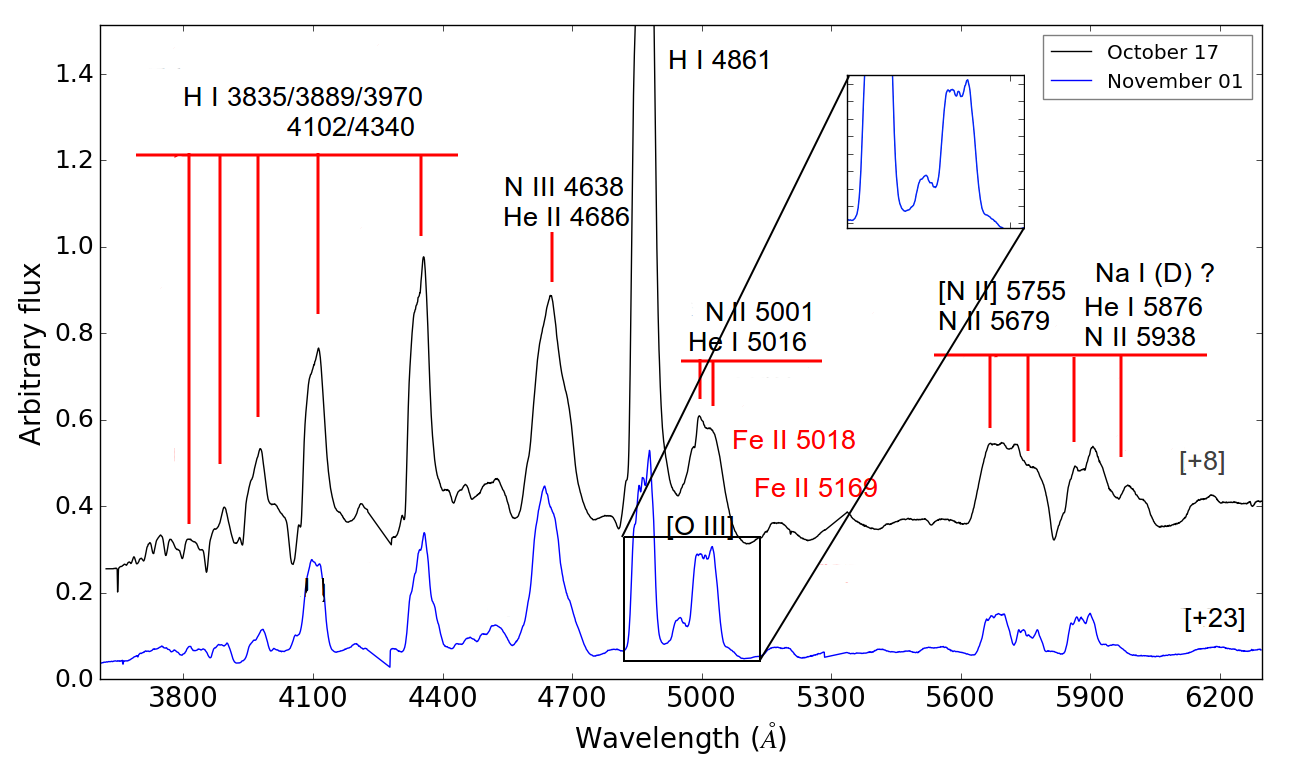}
  \includegraphics[width=\textwidth]{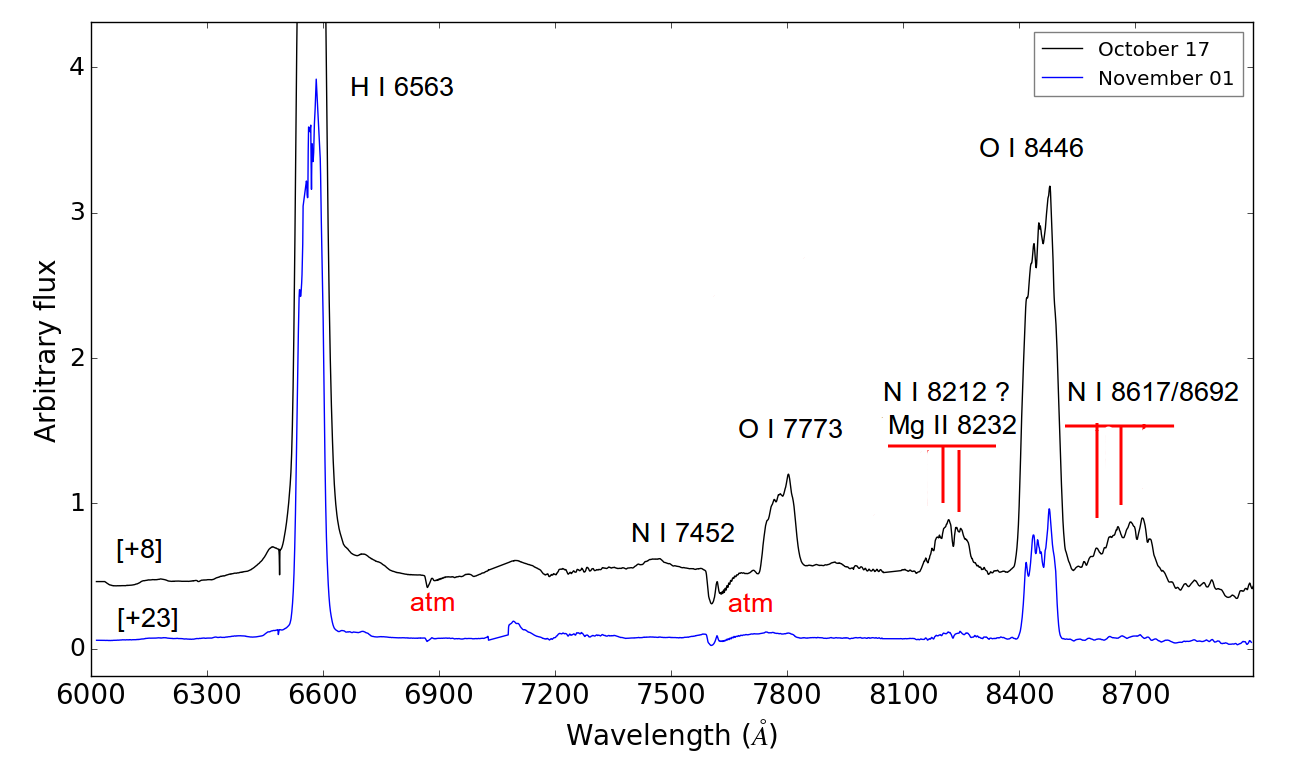}
\caption{The LS medium resolution spectra  plotted between 3600\,--\,6300\,$\mathrm{\AA}$ and 6000\,--\,9000\,$\mathrm{\AA}$ with the flux in arbitrary units. For clarity, the spectra are shifted vertically. The numbers between brackets are days since $t_0$.}
\label{Fig:RSS}
\end{figure*}

\section{Optical spectroscopic observations and data reduction}
\label{spec}
\subsection{SALT medium resolution spectroscopy}
We observed the nova on 2016 October 17 (day 8) and November 1 (day 23) using the Robert Stobie Spectrograph (RSS; \citealt{Burgh_etal_2003}; \citealt{Kobulnicky_etal_2003}), in Long Slit (LS) mode, mounted on the  Southern African Large Telescope (SALT; \citealt{Buckley_etal_2006}; \citealt{Odonoghue_etal_2006}) situated at the SAAO, Sutherland, South Africa. Both observations consist of two spectral ranges; [3500\,$\mathrm{\AA}$ -- 6300\,$\mathrm{\AA}$] and [6000\,$\mathrm{\AA}$ -- 9000\,$\mathrm{\AA}$]. For the October 17 observations the RSS long-slit mode was used, with a 1.5 arcsec  slit and the PG900 grating, resulting in a resolution of $R\sim850$ for the first range and $R\sim1400$ for the second one. 
For the November 1 observations we used a 1.0 arcsec slit and the PG900 grating, resulting in a resolution of $R \sim 1250$ for the first range and $R\sim2000$ for the second one. 
 For both observations, ThAr and Ne lamp arc spectra were taken immediately after the science frames for the first and second range respectively. 

The spectra were first reduced using the  PySALT pipeline \citep{Crawford_etal_2010}, which involves bias subtraction, cross-talk correction, scattered light removal, bad pixel masking, and flat-fielding. The wavelength calibration, background subtraction, and spectral extraction are done using the IRAF (Image Reduction and Analysis Facility) software \citep{Tody_1986}. The spectra are presented in Figure~\ref{Fig:RSS}.

\subsection{SALT high resolution echelle spectroscopy}
The SALT High Resolution Spectrograph (HRS; \citealt{Barnes_etal_2008}; \citealt{Bramall_etal_2010}; \citealt{Bramall_etal_2012}; \citealt{Crause_etal_2014}), a dual-beam, fibre-fed echelle spectrograph, was used to obtain high resolution observations on the nights of 2016 October 17, November 5, November 10, November 23, and December 10 (respectively day 8, 27, 32, 45, and 65). The observations were taken in the low-resolution (LR) mode of HRS to provide a spectrum covering a spectral range of 3800\,$\mathrm{\AA}$ -- 9000\,$\mathrm{\AA}$ at a resolution of $R \sim 15000$. A weekly set of HRS calibrations, including four ThAr + Ar arc spectra and four spectral flats, was obtained in all the modes (low, medium, and high resolution).

\begin{figure*}
  \includegraphics[width=\textwidth]{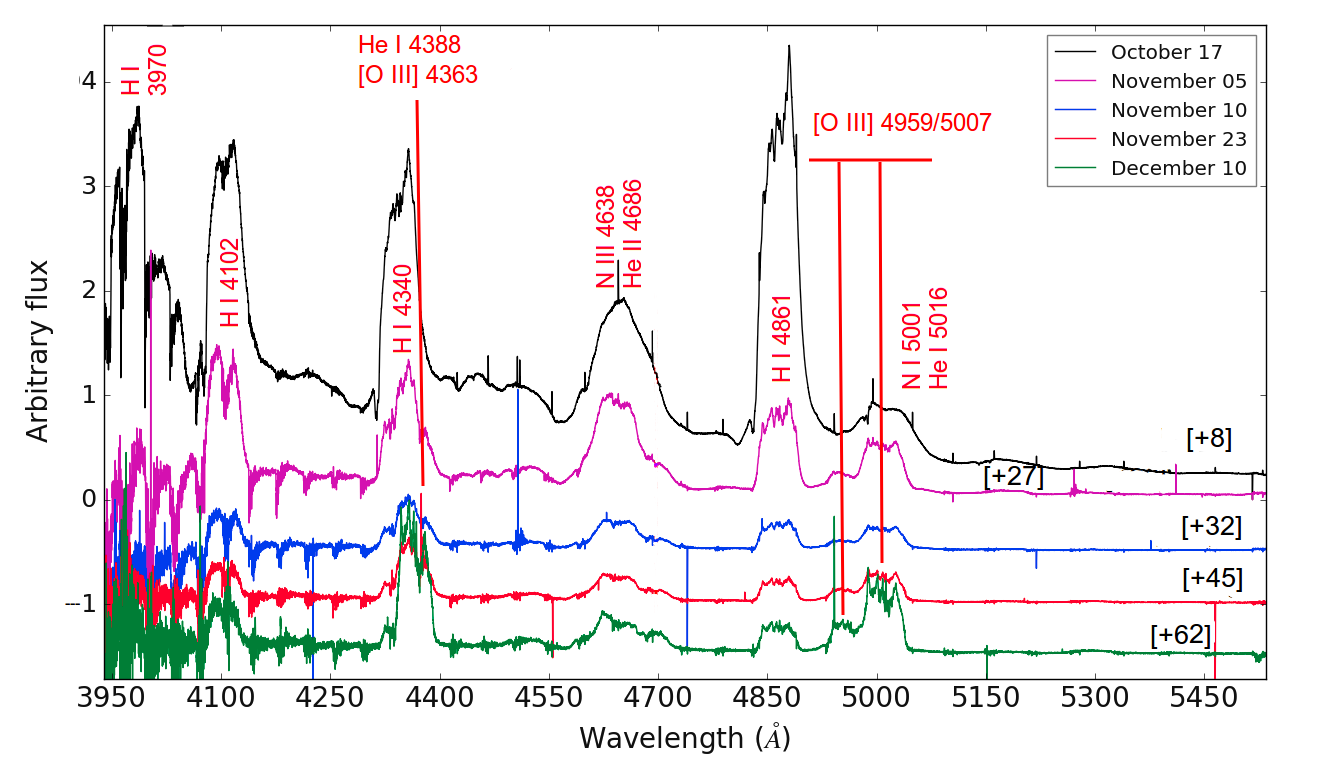}
  \includegraphics[width=\textwidth]{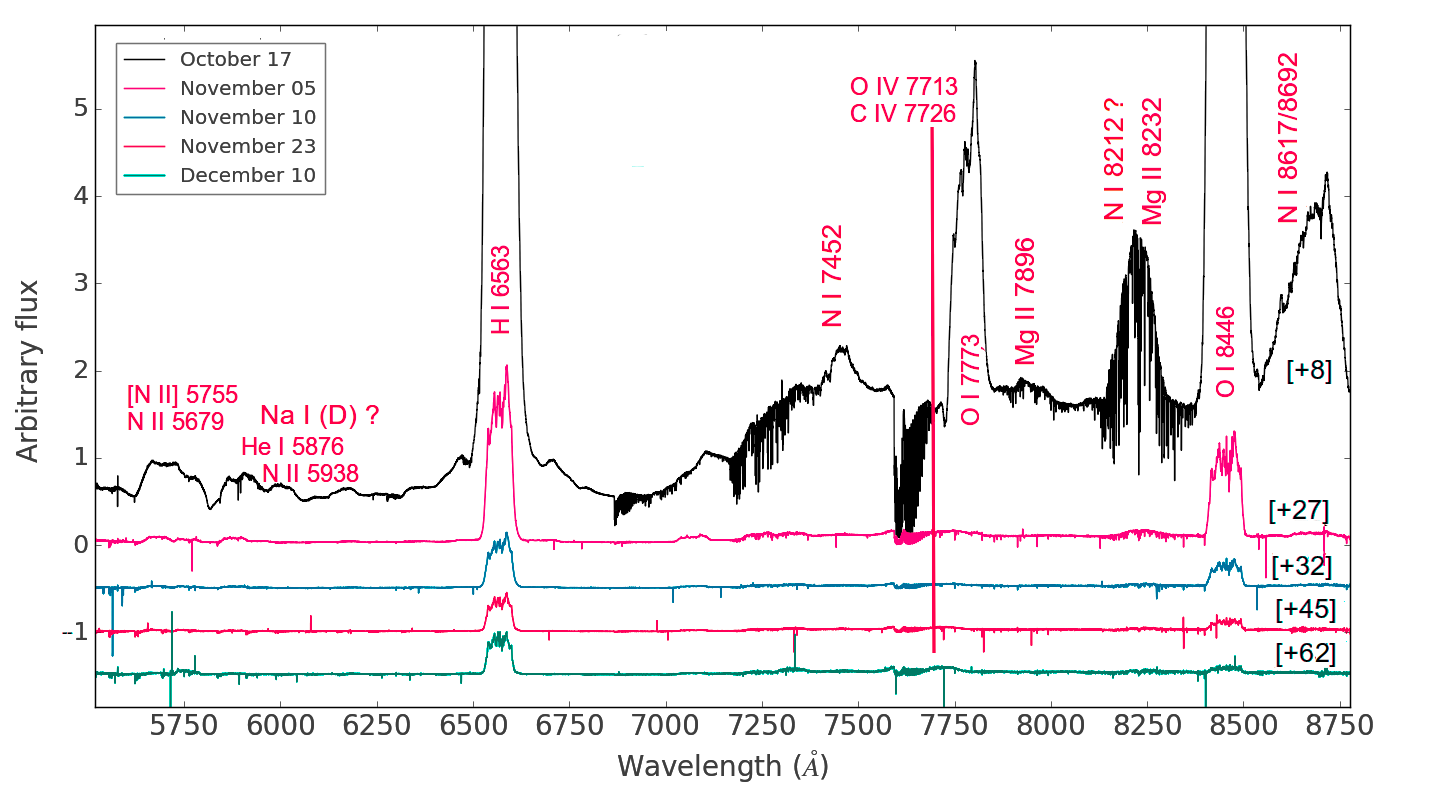}
\caption{The HRS high resolution spectra plotted between 3900\,--\,5500\,$\mathrm{\AA}$ and 5700\,--\,8750\,$\mathrm{\AA}$ with the flux in arbitrary units. For clarity, the spectra are shifted vertically. The numbers between brackets are days since $t_0$.}
\label{Fig:HRSred}
\end{figure*}

 Primary reduction of the HRS data was done using the SALT science pipeline \citep{Crawford_etal_2010}, which includes over-scan correction, bias subtraction, and gain correction. The rest of the reduction was done using the standard MIDAS FEROS \citep{Stahl_etal_1999} and $echelle$ \citep{Ballester_1992} packages. The reduction procedure is described in detail in \citet{kniazev_etal_2016}.  The spectra are presented in Figure~\ref{Fig:HRSred}. 

\subsection{FLOYDS spectroscopy}
\begin{figure*}
  \includegraphics[width=\textwidth]{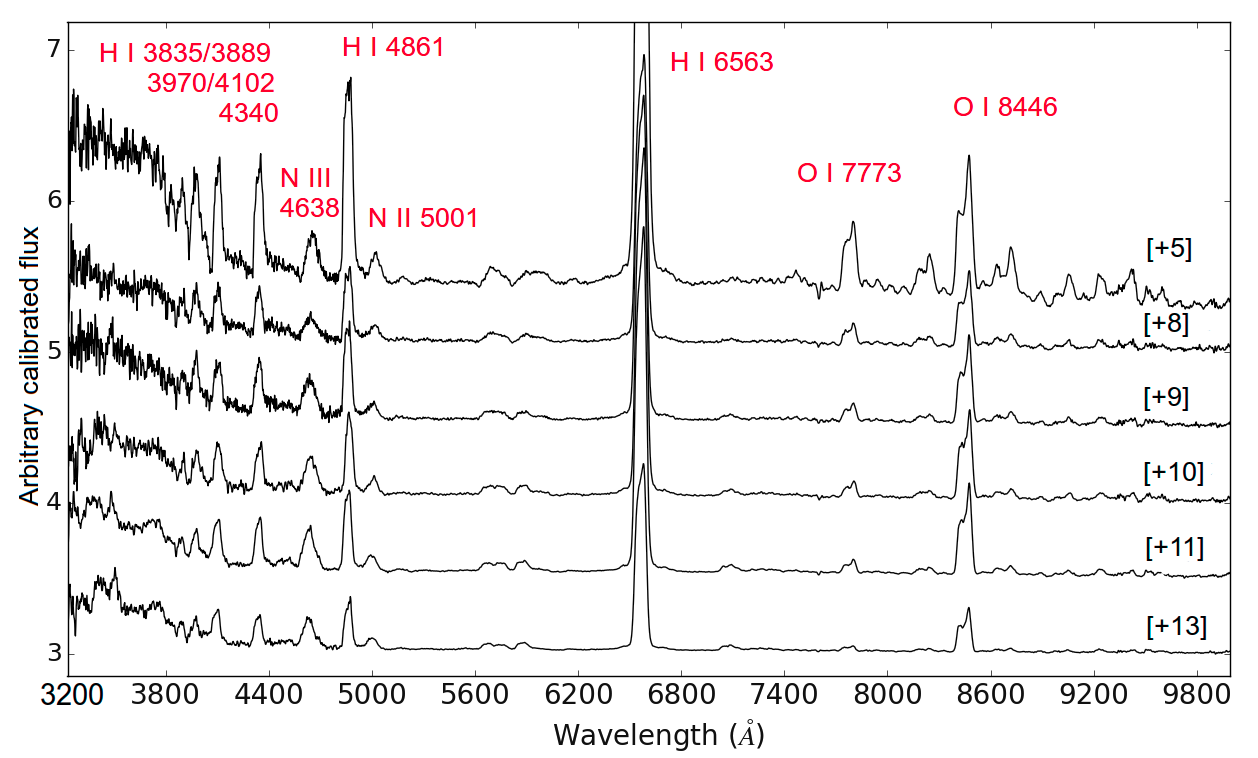}
  \includegraphics[width=1.018\textwidth]{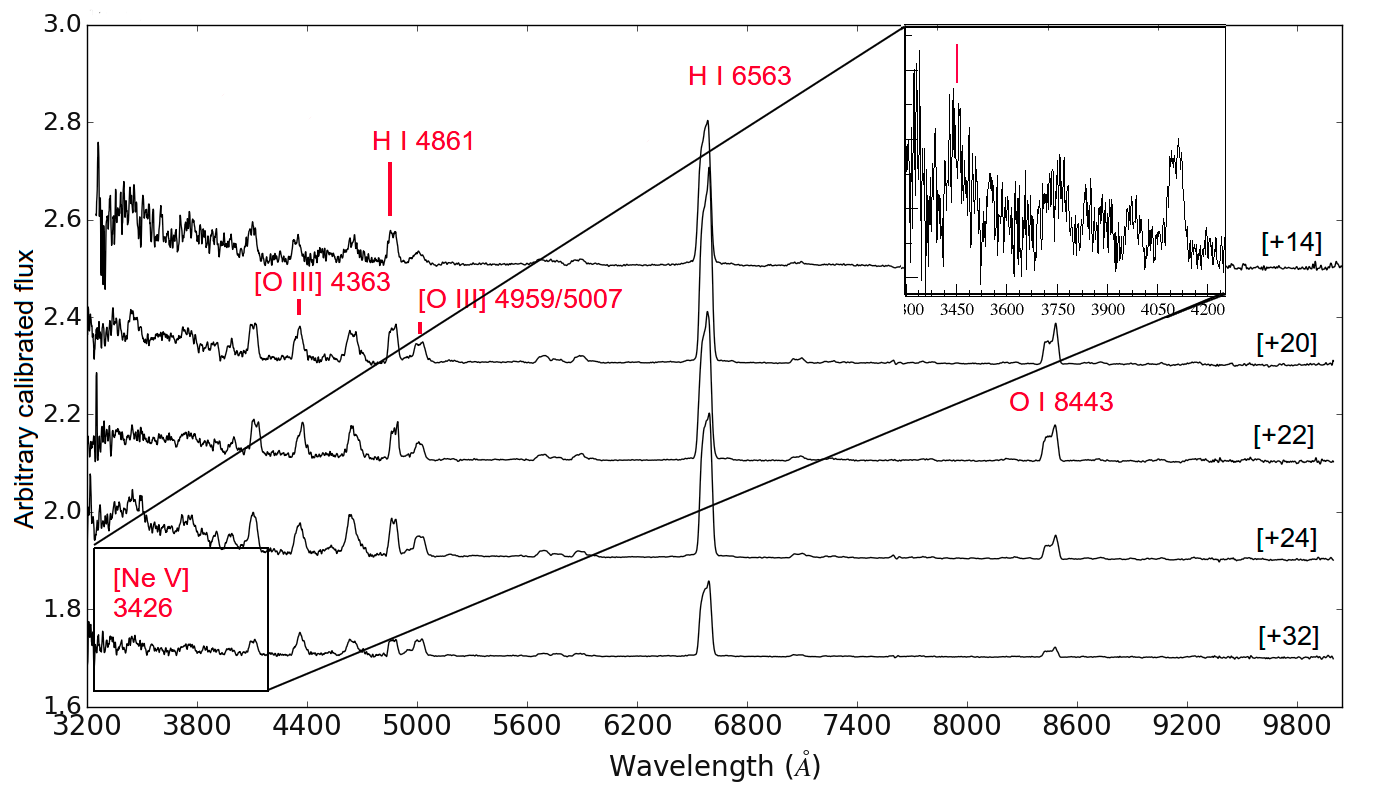}
\caption{The FLOYDS spectra plotted between 3200\,--\,10000\,$\mathrm{\AA}$, with the calibrated flux in arbitrary units. For clarity, the spectra are shifted vertically. From top to bottom: Oct 14, 17, 18, 19, 20, 22, 23, 29,  and 31, Nov 02, and 10. The numbers between brackets are days since $t_0$.}
\label{Fig:fts2}
\end{figure*}

A series of eleven epochs of spectroscopy of nova SMCN 2016-10a were obtained on the nights of 2016 October 14, 17, 18, 19, 20, 22, 23, 29, 31, November 02, and 10 (respectively day 5, 8, 9, 10, 11, 13, 14, 20, 22, 24, 32) using the FLOYDS spectrograph\footnote{\url{https://lco.global/observatory/instruments/floyds}} on the Las Cumbres Observatory \citep[LCO;][]{Brown_etal_2013} 2.0m telescope at Siding Springs, Australia\footnote{Formally known as Faulkes Telescope South.}.  FLOYDS is a cross-dispersed low resolution ($R\sim550$) spectrograph using an Andor Newton DU940P-BU detector and with a wavelength coverage of 3200\,$\mathrm{\AA}$--10000\,$\mathrm{\AA}$. Data reduction was carried out using the FLOYDS data pipeline (v2.2.1).
See Table~\ref{tablefts} for a log of the observations. The flux calibrated spectra are presented in Figure~\ref{Fig:fts2}. Due to poor photometric conditions during the observations as well as slit losses, the absolute flux calibration is not to trust as photometric data.

\subsection{SOAR spectroscopy}
\begin{figure}
  \includegraphics[width=\columnwidth]{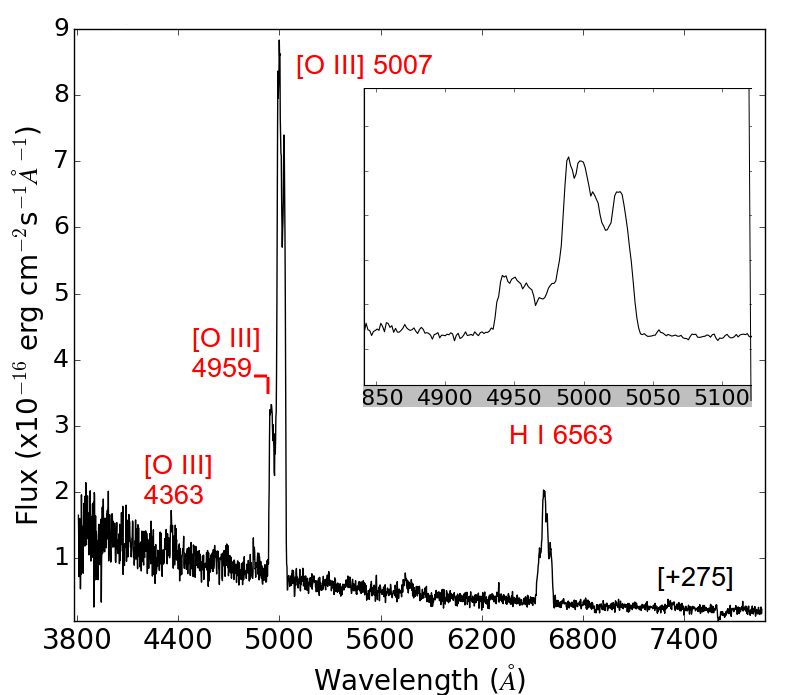}
\caption{The SOAR spectrum of day 275 plotted between 3800\,$\mathrm{\AA}$ -- 7860\,$\mathrm{\AA}$, with the calibrated flux in (erg cm$^{-2}$s$^{-1}$$\mathrm{\AA}^{-1}$). The number between brackets is days since $t_0$. We added a zoom-in plot on the \feal{O}{III} 4959\,$\mathrm{\AA}$ and 5007\,$\mathrm{\AA}$ lines for clarity.}
\label{Fig:SOAR_spec}
\end{figure}

We also performed spectroscopy of the nova on 2017 July 11, using the Goodman spectrograph \citep{Clemens_etal_2004} on the Southern Astrophysical Research (SOAR) telescope. One single 1200\,s spectrum was obtained, using a 400 l mm$^{-1}$ grating and a $1.07\arcsec$ slit, giving resolution of $\sim 7\,\mathrm{\AA}$ over the wavelength range 3810\,$\mathrm{\AA}$ -- 7860\,$\mathrm{\AA}$. The spectrum was reduced in the standard manner, with a relative flux calibration applied, and including an approximate correction for slit losses. The spectrum is presented in Figure~\ref{Fig:SOAR_spec}.

\subsection{Line identification}
\label{spec_lines}
The strongest features in the spectra of day 5 and 8 (Figure~\ref{Fig:RSS},~\ref{Fig:HRSred}, and~\ref{Fig:fts2}) are the broad  Balmer lines (H${\alpha}$, H${\beta}$,  H${\gamma}$,  H${\delta}$, H${\epsilon}$, H${\zeta}$, and H${\eta}$). The hydrogen Balmer lines are accompanied by weak absorption features. The radial velocities (RVs) derived from the P Cygni absorption on day 8, by measuring the wavelengths at the minimum flux, are as follows: H${\gamma}$\,$\sim$\,$- 2300\,\pm200$\,km\,s$^{-1}$, H${\delta}$\,$\sim$\,$- 2700\,\pm200$\,km\,s$^{-1}$, H${\epsilon}$\,$\sim$\,$- 2700\,\pm200$\,km\,s$^{-1}$, and H${\zeta}$\,$\sim$\,$- 2500\,\pm200$\,km\,s$^{-1}$. Another set of prominent features are the Oxygen lines (\eal{O}{I} 7773 and 8446\,$\mathrm{\AA}$). Nitrogen emission lines are also present (\eal{N}{III} 4638\,$\mathrm{\AA}$, \eal{N}{II} 5001\,$\mathrm{\AA}$, 5679\,$\mathrm{\AA}$, 5938\,$\mathrm{\AA}$, \feal{N}{II} 5755\,$\mathrm{\AA}$, \eal{N}{I} 7452\,$\mathrm{\AA}$, 8212\,$\mathrm{\AA}$, 8617\,$\mathrm{\AA}$, and 8692\,$\mathrm{\AA}$), noting that the \eal{N}{III} 4638\,$\mathrm{\AA}$ might be accompanied by a weak P Cygni absorption. There is a possible weak detection of \eal{He}{I} 7065\,$\mathrm{\AA}$. \eal{He}{II} 4686\,$\mathrm{\AA}$ and \eal{He}{I} 5016\,$\mathrm{\AA}$ might be blended with their neighbouring nitrogen lines. \eal{He}{I} 5876\,$\mathrm{\AA}$ might also be affected by \eal{Na}{I}D and the nitrogen line at 5938\,$\mathrm{\AA}$. \eal{Mg}{II} 7896\,$\mathrm{\AA}$ and 8232\,$\mathrm{\AA}$ are also present. The line at 5169\,$\mathrm{\AA}$ is possibly \eal{Fe}{II} while the other multiplet (42) lines at 4924\,$\mathrm{\AA}$ and 5018\,$\mathrm{\AA}$ might be blended with other lines. 

The spectra from day 9 to 14 show the same emission features with the \eal{O}{I} 7773\,$\mathrm{\AA}$ and \eal{N}{II} 5001\,$\mathrm{\AA}$ lines fading gradually. The spectra of day 20 to 23 (Figure~\ref{Fig:RSS}, and~\ref{Fig:fts2}) show similar lines, with the appearance of \feal{O}{III} 4959\,$\mathrm{\AA}$ and 5007\,$\mathrm{\AA}$ nebular lines. The weak absorption features had faded. The \eal{O}{I} 7773\,$\mathrm{\AA}$ line weakened compared to the 8446\,$\mathrm{\AA}$ line. The \feal{N}{II} 6548\,$\mathrm{\AA}$, 6583\,$\mathrm{\AA}$ might be blended with the broad H${\alpha}$ line (see section~\ref{Balmer}). 
\\

In the spectra of day 32 (Figure~\ref{Fig:fts2}) forbidden Ne lines emerge at 3346\,$\mathrm{\AA}$ and 3426\,$\mathrm{\AA}$ (see Section~\ref{UV_spec}), while the Balmer features start to fade compared to the \feal{O}{III} emission lines (4363\,$\mathrm{\AA}$, 4959\,$\mathrm{\AA}$, 5007\,$\mathrm{\AA}$). The \eal{O}{I} lines at 7773\,$\mathrm{\AA}$ and 8446\,$\mathrm{\AA}$ further in the red also faded noticeably. The spectra of day 45 and 65 show strong emission lines of \feal{O}{III} and \eal{He}{I} 4388\,$\mathrm{\AA}$ in the blue while the \eal{O}{IV} 7713\,$\mathrm{\AA}$ and \eal{C}{IV} 7726\,$\mathrm{\AA}$ nebular lines emerge.

The late spectra of day 275 (Figure~\ref{Fig:SOAR_spec}) were dominated by the \feal{O}{III} 4363\,$\mathrm{\AA}$, 4959\,$\mathrm{\AA}$, and 5007\,$\mathrm{\AA}$ lines, along with H$\alpha$. The spectral line identification was primarily done using the list from \citet{Williams_2012} and the Multiplet Table of Astrophysical Interest \citep{Moore_1945}. In Table~\ref{line_det} we list the line IDs along with the FWHM, Equivalent Widths (EWs), and fluxes for those emission lines for which an estimate was possible.

\subsection{SALT spectropolarimetry}
\label{spectropol}
Spectropolarimetry observations were obtained on the nights of 2016 October 24 and November 21 (days 15 and 43) using the RSS spectrograph. The RSS is capable of imaging polarimetry and spectropolarimetry (for more details  on the optics see \citealt{Nordsieck_etal_2003}). Both observations consist of two spectral ranges; [3500\,$\mathrm{\AA}$ -- 6300\,$\mathrm{\AA}$] and [6000\,$\mathrm{\AA}$ -- 9000\,$\mathrm{\AA}$]. The RSS LS mode was used, with a 1.5 arcsec slit and the PG900 grating, resulting in a resolution of $R\sim850$ for the first range and $R\sim1400$ for the second one.  ThAr and Ne lamp arc spectra were taken immediately before the science frames for the first and second range, respectively.
\\

The spectropolarimetry data reduction was carried out using the PolSALT pipeline (Crawford et al. in preparation). This pipeline performs basic reductions (e.g. bias subtraction, cross-talk correction, over-scan correction, and cosmic-ray rejections), wavelength calibration and spectrum extraction, whereafter the Stokes parameters are determined.
\\

\begin{figure*}
  \includegraphics[width=\textwidth]{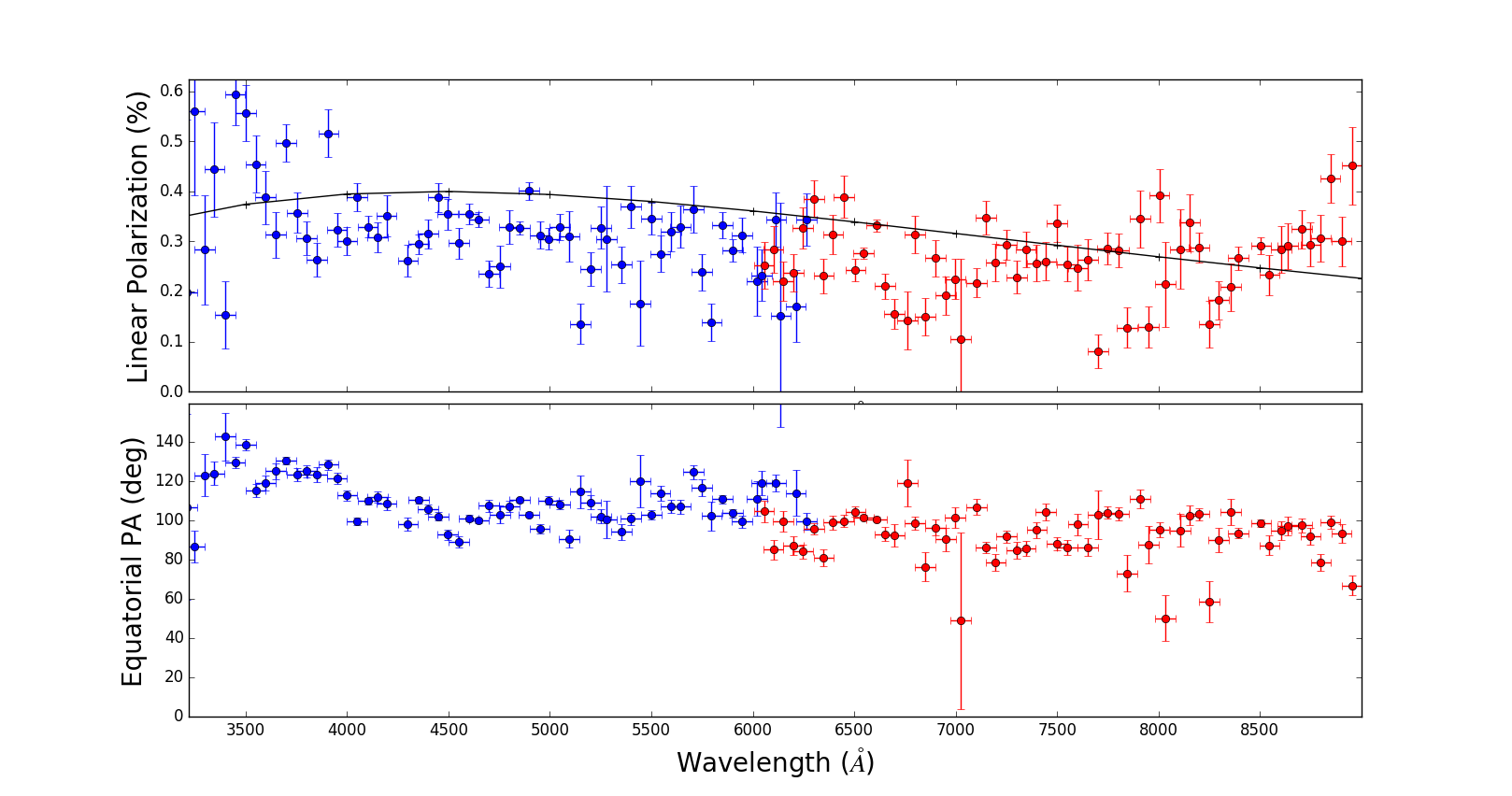}
\caption{The linear polarization (top) and position angle (bottom) plotted against the wavelength from day 15. The blue and red points correspond to observations from the first and second spectral range respectively. The black solid line represents the values derived from the Serkowski relation, assuming $p_{\mathrm{max}}$ = 0.4\% at $\lambda$ = 4450\,$\mathrm{\AA}$, for the polarization due to the interstellar medium. A colour version of this plot is present in the online journal.}
\label{Fig:spectropol}
\end{figure*}

The linear polarization, and the position angle are plotted as a function of wavelength in Figure~\ref{Fig:spectropol}. We show the observations from day 15 as a sample for both observations. The enhancement at the blue and red ends of the spectra might be due to low signal to noise ratio and hence we consider only the range between 4000\,$\mathrm{\AA}$ and 8000\,$\mathrm{\AA}$ in our analysis. The linear polarization is on average $\sim$ 0.3\%. This may be due to the interstellar medium or intrinsic to the nova, or both. \citet{Rodrigues_etal_1997} studied the interstellar polarization toward the SMC. Nova SMCN 2016-10a is situated just south of their region III which has the smallest foreground polarization ($P$ = 0.17\%).

In order to estimate the contribution of the interstellar medium, we derive the linear polarization at different wavelengths using the Serkowski empirical relation \citep{Serkowski_etal_1975,Wilking_etal_1980}, and assuming $p_{\mathrm{max}}$ = 0.4\% at $\lambda$ = 4450\,$\mathrm{\AA}$. The values are plotted in Figure~\ref{Fig:spectropol}. The polarization is below the values derived for the interstellar medium.

Although, the interstellar polarization is low, it is nevertheless likely to be the primary source of polarization in our observations. There is no sign of de-polarization effect due to emission lines that is usually observed in case of polarization originating from novae (see e.g. \citealt{Ikeda_etal_2000}). The polarization in novae can originate either from scattering on dust or from a magnetic field, so if the observed polarization is indeed from the interstellar medium, the lack of intrinsic polarization can indicate the following: First, that there is no dust in the ejecta, consistent with the behaviour of the optical and infrared light-curves. Secondly, there might be no or weak magnetic field.

\section{X-ray observations}
\label{xray}
{\em Swift} \citep{Gehrels_etal_2004} observations of nova SMCN 2016-10a started on 2016 October 15 (6.1
days after $t_0$). At this time, no X-rays were detected, but a
500\,s grism spectrum showed a bright UV source with strong lines \citep{ATel_9635}. Additional observations were performed on the following two
days, after which the cadence was lowered to once every few days. On 2016 November
07 (day 28.2), a bright, soft X-ray source was detected, at a count rate of
1.57~$\pm$~0.06\,count\,s$^{-1}$. In order to parameterize the onset of the SSS phase, multiple short snapshots of data were
collected on 2016 November 11-12 (days 33-34), revealing the high amplitude flux
variability seen in a number of other novae (e.g. \citealt{Ness_etal_2009,
Osborne_etal_2011,Bode_etal_2016}). Observations then continued approximately every two to four days until 2017 May 12, with further observations following an eight day cadance running until 2017 June 25. Five further observations were obtained
between 2017 July 8 and August 24. We note that the location of the SMC is not observable by {\em Swift} for up to 10 days each month because of the so-called ``pole constraint''; this arises from the requirement that {\em Swift} not point too close to the Earth limb.

The {\em Swift} data were processed with HEASoft 6.19, and analyzed with
the most up-to-date calibration files. The X-ray Telescope (XRT; \citealt{Burrows_etal_2005}) was operated in ``Auto"
state, so that the most appropriate mode (Windowed Timing - WT - when the
source was above about 5\,count\,s$^{-1}$; Photon Counting - PC - when the
source was fainter) would be automatically chosen. Reaching a peak count
rate of $\sim$~50\,count\,s$^{-1}$, the WT data did not suffer from pile-up,
so a circular extraction region was used for these data. PC mode data
above around 0.5\,count\,s$^{-1}$ were considered to be affected by pile-up,
so, at these times, the core of the Point Spread Function was excluded,
and corrected for, as required, using annular extraction regions.
Background counts were estimated from nearby, source-free regions. In order to facilitate Cash statistic \citep{Cash_1979} fitting within {\sc xspec} \citep{Arnaud_1996}, the X-ray spectra were binned to ensure a minimum of 1 count\,bin$^{-1}$. Fitting using the C-statistic (modified by {\sc xspec} to work in the case of a background spectrum) provides less biased parameter estimates, even in the high-count regime \citep{Humphrey_etal_2009}.
\begin{figure}
\begin{center}
  \includegraphics[width=\columnwidth]{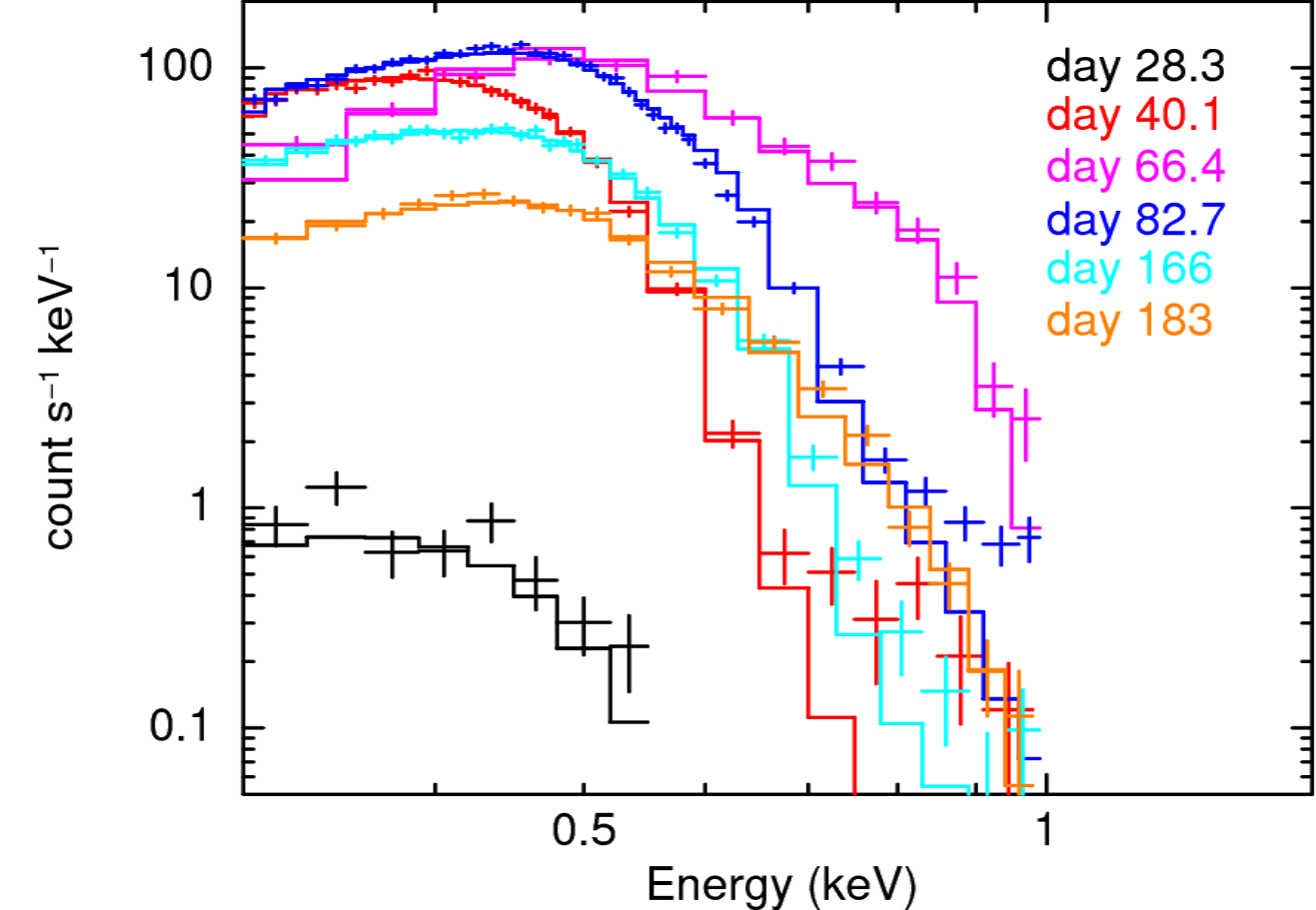}
\caption{A sample of {\em Swift} X-ray spectra from days 28.3, 40.1, 66.4, 82.7, 166 and 183 since eruption ($t_0$), fitted in each case with the best-fit TMAP
atmosphere model (see Figure~\ref{Fig:NovaSMC2016_atmos}). A colour version of this plot is present in the online journal.}
\label{Fig:papercomp}
\end{center}
\end{figure}

\begin{figure*}
\begin{center}
  \includegraphics[width=\textwidth]{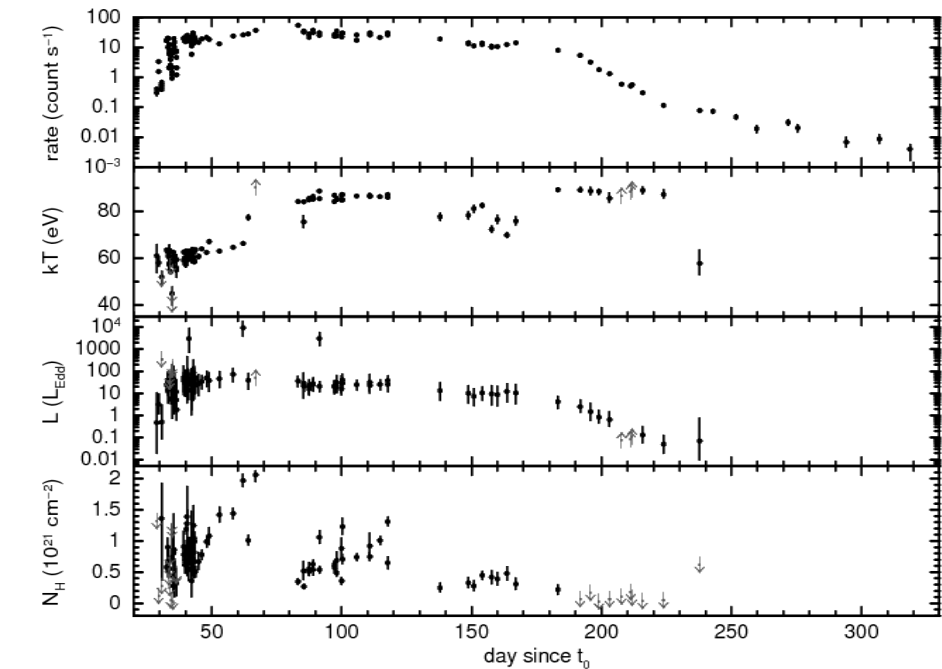}
\caption{From top to bottom: the count rate, the temperature of the SSS emission, the bolometric luminosity in units of the Eddington value for a 1.2\,M$_{\odot}$ WD, and the absorbing column density plotted against the days since $t_0$ (HJD 2457670.70). All parameters were determined from the TMAP model atmosphere fits as described in the text. Grey arrows indicate upper or lower limits (90 \%) as necessary. For the luminosity calculation, a distance = 61\,kpc is used.}
\label{Fig:NovaSMC2016_atmos}
\end{center}
\end{figure*}

\begin{figure*}
\begin{center}
  \includegraphics[width=\textwidth]{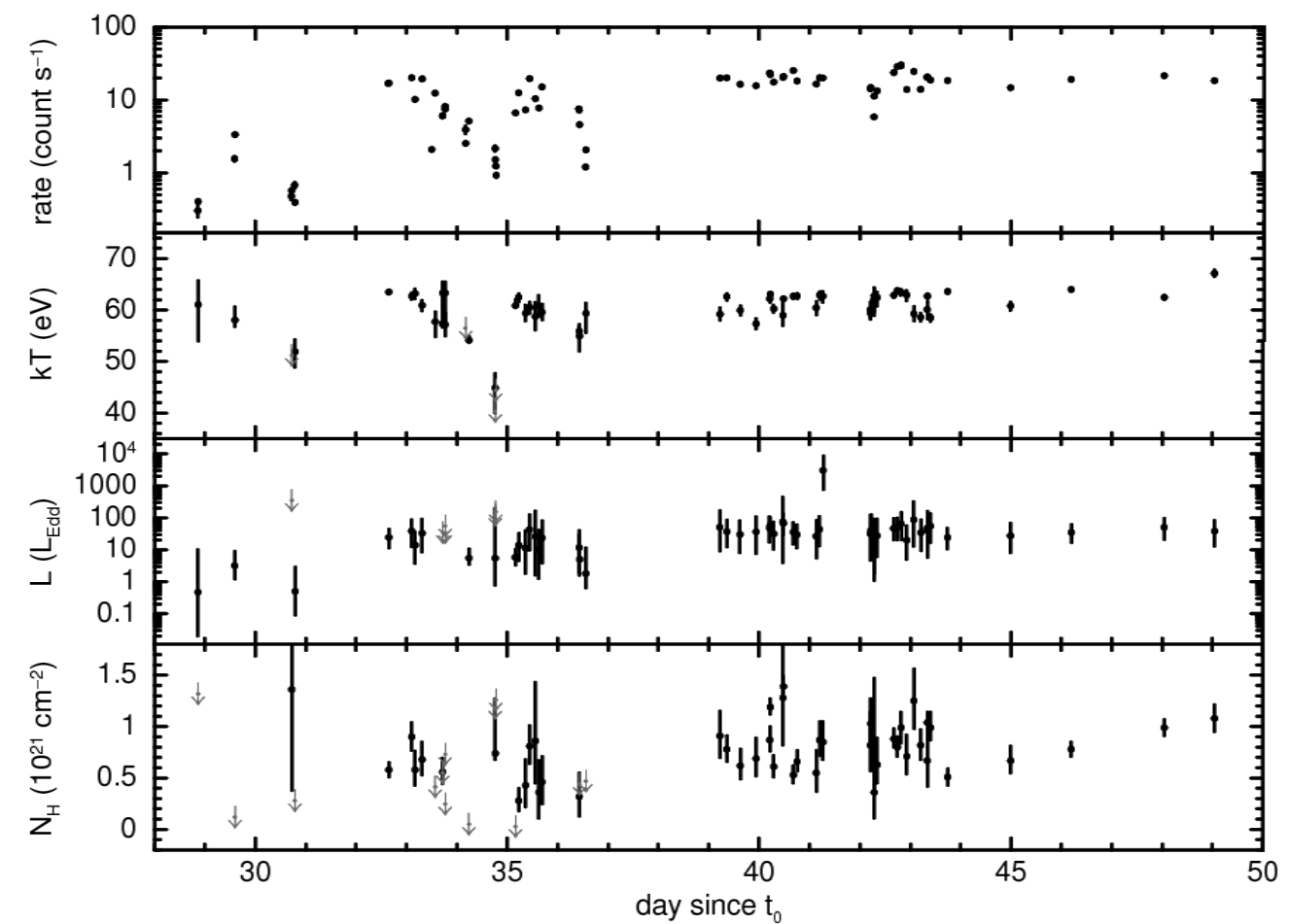}
\caption{As for Figure~\ref{Fig:NovaSMC2016_atmos}, but a zoomed-in version for the observations up to day 50 showing the early variability more clearly.}
\label{Fig:10}
\end{center}
\end{figure*}

The first X-ray detection of nova SMCN 2016-10a showed the emission already to be super-soft. At the distance of the SMC, any underlying shock emission would be likely to be faint and difficult to detect (cf. Nova LMC 2009a;  \citealt{Bode_etal_2016}). Assuming optically thin emission at a temperature of 5\,keV, and combining all the X-ray data collected, a 90\% upper limit of $\sim2\times10^{34}$\,erg\,s$^{-1}$ can be placed on the unabsorbed  0.3-10\,keV luminosity for any hard spectral component. The X-ray spectra were therefore fitted below 1\,keV with an absorbed plane-parallel,
static, non-local-thermal-equilibrium stellar atmosphere model (grid 003)\footnote{TMAP: T{\" u}bingen NLTE Model Atmosphere Package:
http://astro.uni-tuebingen.de/\raisebox{.2em}{\tiny$\sim$}rauch/TMAF/flux\_HHeCNONeMgSiS\_gen.html};
samples of the spectra are shown in Figure~\ref{Fig:papercomp}, and the
parameter results from the fitting are shown in
Figure~\ref{Fig:NovaSMC2016_atmos} while Figure~\ref{Fig:10} focuses in on the early variability phase. This highlights the order of magnitude or more variability seen in the X ray count rate over an interval of two days as the SSS became established. The atmosphere grids typically provide good fits, with the reduced Cash statistic values mainly lying in the range 0.8-1.2. The model atmosphere is limited to temperatures of between $\sim$~38\,eV and 91\,eV, and a small number of the spectra for this nova had temperatures which hit the upper or lower limits of these bounds, suggesting the actual values are outside the range covered by the grids; these are shown by upper
or lower limits in the second panel of Figure~\ref{Fig:NovaSMC2016_atmos} and Figure~\ref{Fig:10} as
necessary. The third panel of Figure~\ref{Fig:NovaSMC2016_atmos} and~\ref{Fig:10} show the
bolometric luminosity of the soft emission component, in terms of the
Eddington luminosity for a 1.2\,M$_{\odot}$ WD, assuming a distance of 61\,kpc. The absorbing column was
allowed to vary in the fits, and is shown in the bottom panel of
Figure~\ref{Fig:NovaSMC2016_atmos} and~\ref{Fig:10}.

The temperature of the SSS was lower at earlier times, increasing suddenly after day 60, and reaching an apparent peak (as shown by the magenta spectrum in Figure~\ref{Fig:papercomp}) at a time
the source location unfortunately became non-observable by {\em Swift}. Following the emergence from the observing constraint (day 82.7), the temperature of the SSS emission was consistently high, $\sim$ 90 eV.  We note that there are occasional fast changes in the X-ray count rate, for example around day 36.5, without corresponding variations in the spectral shape (i.e. the temperature remains approximately constant). This implies that there must be partial obscuration by clumps of high optical depth material at these times.

Even as the X-ray emission started to fade sharply, after day 180, the temperature remained high, actually increasing slightly compared to earlier spectra. However, on day 237 a significant drop in temperature was observed. A final cooling of the X-ray emission is expected \citep{Soraisam_etal_2016}, and has been seen in other novae well-monitored by {\em Swift} \citep{Page_Osborne_2014}. From day 242 onwards, the X-ray spectrum was not well fitted by the atmosphere model, with the shape better parametrized by an optically thin component with kT $\sim$ 80\,eV. These data-sets have therefore not been included in panels 2-4 in Figure~\ref{Fig:NovaSMC2016_atmos}. This adds to the evidence that the eruption has ended. As the X-rays faded, there was no longer any evidence for excess absorption above the interstellar value in the direction of the SMC  (6$\times$10$^{20}$ cm$^{-2}$; \citealt{Coe_etal_2011}); in fact, some of the late-time upper limits suggest that the total absorbing column is below this value suggesting either an uncertainty in the absorption in the direction of the SMC, or a limitation of the emission model applied to the X-ray spectra.

\citet{Krautter_etal_1996} first showed that the X-ray emission from a fading post-eruption nova remains soft. The residual material is hot and extended and, even as the nuclear fusion comes to an end, the evolving WD decreases in radius (towards the quiescent radius of the star) at nearly the same luminosity for a while, leading to an increase in temperature.

A number of novae with bright X-ray emission have shown short period ($<$
100\,s) oscillations (e.g. \citealt{Osborne_etal_2011,ATel_2423,Ness_etal_2015}). We have therefore searched for the presence of such
oscillations in the soft X-ray light-curves of nova SMCN 2016-10a, following
the procedure outlined in Beardmore et. al. (in prep.).

WT light-curves were extracted with a time bin size of 17.8\,ms over the
energy range 0.3-1.0\,keV (i.e. where the SSS emission is dominant).
Periodograms were then computed from continuous light-curve sections up to
512\,s duration -- given the short nature of some of the XRT snapshots,
those with exposures less than 256\,s were rejected while those between
256--512\,s were padded to a duration of 512\,s using the mean rate for the
interval. The average periodogram constructed from 80 individual ones spanning days 32.03 to 157.07 reveals no evidence for any oscillations with a timescale ranging
from 1 - 256\,s, with a 3 sigma fractional rms upper limit of 1.4 per
cent.

\section{UV observations}
\label{UV}
\subsection{{\em Swift} UVOT photometry}

\begin{figure*}
\begin{center}
  \includegraphics[width=\textwidth]{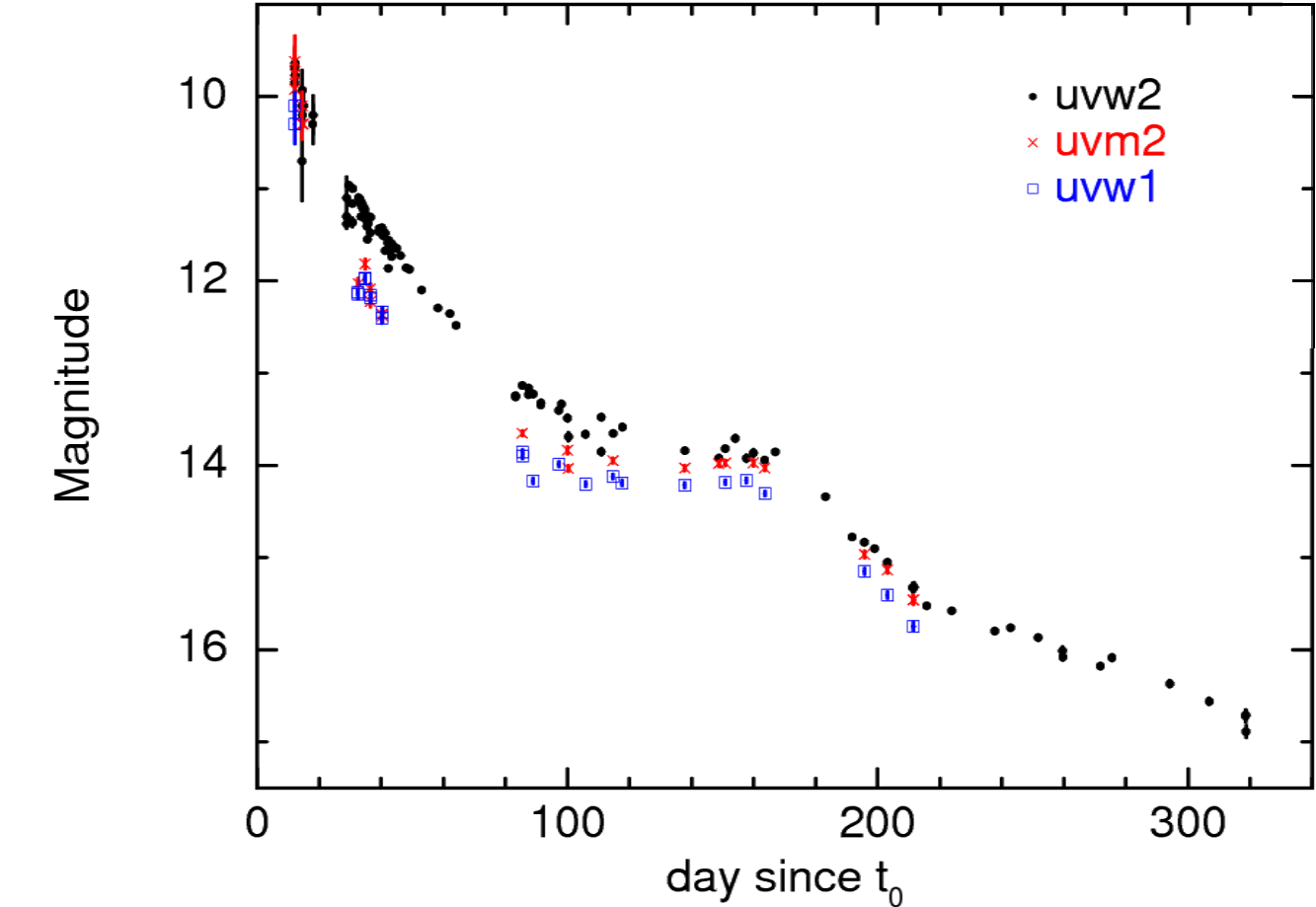}
\caption{The UVOT light-curve of nova SMCN 2016-10a plotted against time in days since $t_0$.}
\label{Fig:uvot_lightcurve}
\end{center}
\end{figure*}

UV/Optical Telescope (UVOT; \citealt{Roming_etal_2005}) photometric observations started on day 12 (2016 October 21) and continued until day 319 (2017 August 24). The initial photometry was obtained using the readout streak method \citep{Page_etal_2013}, as the nova was too bright for normal aperture photometry, described in \citet{Poole_etal_2008}. Each image was inspected and those where the spacecraft failed to achieve a lock-on were discarded. The observations were mainly in the uvw2 band, with additional observations in the UVOT uvm2 and uvw1 bands, with central wavelengths: $uvw2$ = 1928\,$\mathrm{\AA}$; $uvm2$ =
2246\,$\mathrm{\AA}$; $uvw1$ = 2600 \,$\mathrm{\AA}$. Figure~\ref{Fig:uvot_lightcurve} illustrates  the UVOT light-curve. There is a plateau in the UV emission between at least days 90 and 170,
approximately coinciding with the main interval of bright, high-temperature, super-soft X-rays
(Figure~\ref{Fig:NovaSMC2016_atmos}). While it is hard to be definitive, due both to data gaps and small scale variability, the plateau in the uvw2 filter may start slightly later and end slightly earlier than the soft X-ray plateau. Such a plateau phase has been seen in other novae (e.g. \citealt{Kato_etal_2008,Page_2013}). Note that the SMARTS optical photometry shows a plateau as well, starting about day $\sim$ 100 (consistent with $uvw1$), and ending after day $\sim$ 130, during solar conjunction (see Figure~\ref{Fig:BVRI_LC}).

We looked for temporal structure in the uvw2 photometry by computing the n = 2, 3, and 
4 structure functions using the method described in \citet{Saxton_etal_2012}. 
No evidence of a preferred timescale in the range of 0.01-10 d was found.

\subsection{{\em Swift} UVOT spectroscopy}
\label{UV_spec}
The initial {\em Swift} UVOT observations, which were obtained on 2016 October 15, 
6.1 days after the eruption, used the UV grism. On the next two days, two further UV spectra were obtained \citep{ATel_9635}. The nova brightness was just below the limit where the flux calibration is possible \citep{Kuin_etal_2015}. The initial {\em Swift} Target of Opportunity observations were taken with no offset, slew in place, or photometry, to allow for quick scheduling of the observation. 

Starting on 2016 October 21, we continued with UV grism clocked-mode \citep{Kuin_etal_2015} observations every two days  as scheduling allowed. These were typically taken at an offset of 5.6\arcmin vertically in the detector plane so that the overlap of zeroth orders of field stars is limited to wavelengths shorter than about 2100\,\AA. For the periods 2016 November 7-14 and 2016 December 31 to 2017 January 6, the observations were made at no offset, and during the period of 2016 November 23 to December 12, the offset varied with the roll angle. The main effect is that in the spectra with no offset there are clear signatures of second order emission lines present, namely those at 1750\,$\mathrm{\AA}$, 1909\,\,$\mathrm{\AA}$, and 2143\,\,$\mathrm{\AA}$ in second order appear around 2880\,$\mathrm{\AA}$, 3310\,$\mathrm{\AA}$, and 3910\,$\mathrm{\AA}$  in the first order. These emission lines have also  been found in other CO novae, like nova V339 Del \citep{ATel_5967,Shore_etal_2016}. The line at 3130\,$\mathrm{\AA}$ is likely the \eal{O}{III} 3133.7\,$\mathrm{\AA}$ resonance line \citep{Young_etal_2011} and is being fluoresced by the \eal{He}{II} 304 EUV line. For that reason it is not affected by optical thickness effects as the other resonance lines are. The \eal{O}{II} 2471\,$\mathrm{\AA}$ line is probably present, but weak and blended with other weak features.  

\begin{figure*}
  \includegraphics[width=\textwidth]{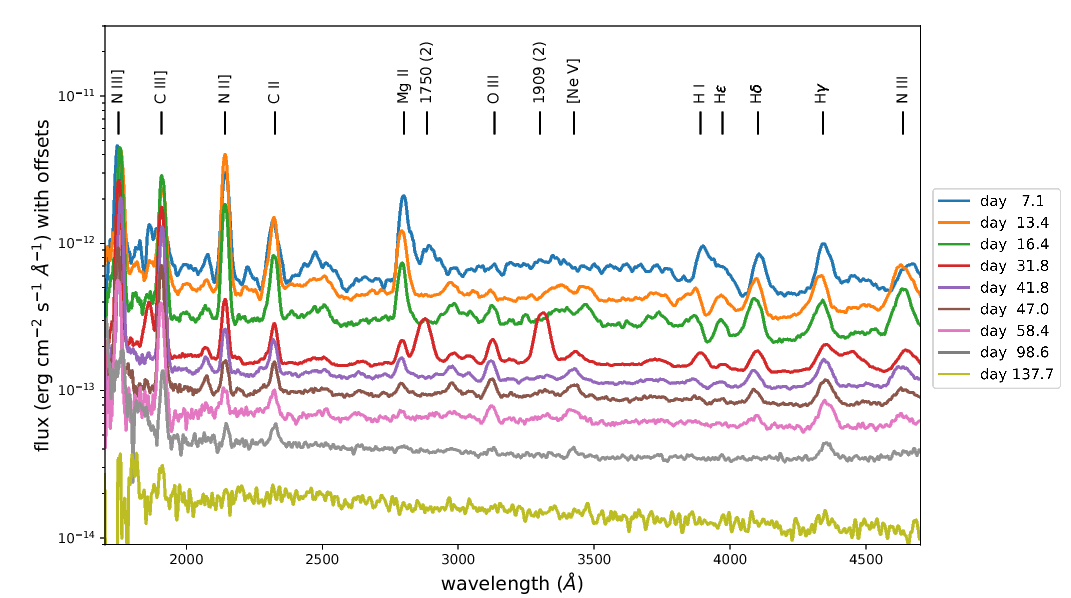}
\caption{The UV grism summed spectra. For clarity the spectra are all shifted vertically, starting with an offset of 1e-14 then incrementally by 2e-14 in flux. The numbers to the right are days since eruption ($t_0$). A colour version of this plot is present in the online journal.}
\label{Fig:uvgrism_spectra}
\end{figure*}

The spectra were extracted from each image using the {\tt UVOTPY} software \citep{Kuin_2014} with the calibration of \citet{Kuin_etal_2015}. The grism images were inspected to identify contamination from zeroth and first orders of other sources. The spectra were summed in groups of 3 to 5 exposures to improve the continuum which retains a noise contribution from irreducible MOD-8 noise \citep{Kuin_etal_2015}. Prior to summing, the individual spectra were lined up by shifting them in pixel/bin space, re-computing the wavelengths after the shift. This is needed as the dispersion changes from the short to long wavelengths. The dates of the summed spectra are provided in Table~\ref{uvgrismtable}. The spectra are illustrated in Figure~\ref{Fig:uvgrism_spectra}. Up to day 9 there is unresolved absorption in the 2550-2650\,$\mathrm{\AA}$ range consistent with the Fe-curtain (optically thick) still being weakly present. The UV spectra show strong emission from \hfeal{N}{III} 1750\,$\mathrm{\AA}$, \hfeal{C}{III} 1909\,$\mathrm{\AA}$, \eal{N}{II} 2143\,$\mathrm{\AA}$, \eal{C}{II} 2324\,$\mathrm{\AA}$, \eal{Mg}{II} 2800\,$\mathrm{\AA}$, and \feal{Ne}{III} 3868\,$\mathrm{\AA}$, 3967\,$\mathrm{\AA}$. Up to day 43 we see the \eal{Al}{III} 1860\,$\mathrm{\AA}$ resonance line, while the lines from \eal{O}{III} 3130\,$\mathrm{\AA}$ and \feal{Ne}{V} 3346\,$\mathrm{\AA}$, 3426\,$\mathrm{\AA}$ become prominent after day 28, suggesting a gradual increase in ionization. After day 117 post-eruption, the UV grism data (4\,ks summed exposure) become too noisy showing only evidence of the continued presence of the \hfeal{N}{III} 1750\,$\mathrm{\AA}$  and \hfeal{C}{III} 1909\,$\mathrm{\AA}$ lines.

\begin{figure*}
  \includegraphics[width=\textwidth]{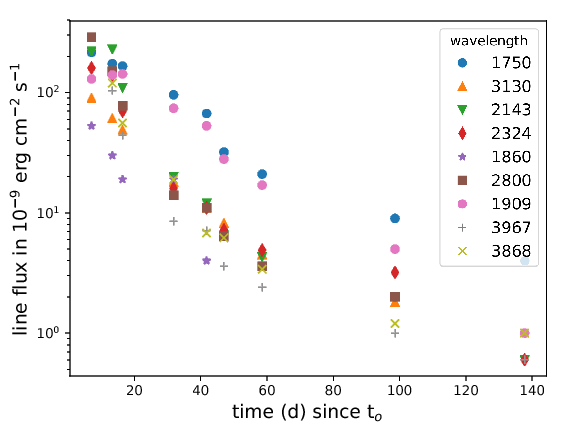}
\caption{The evolution with time post-eruption of the line fluxes in the UVOT spectra, colour and symbol coded based on the lines rest-wavelength.}
\label{Fig:UV_line_flux}
\end{figure*}

The line flux was determined by integrating the flux above the nearby continuum. Some lines were difficult to measure. The \feal{Ne}{V} 3346\,$\mathrm{\AA}$ line suffers from the nearby \hfeal{C}{III} 1909\,$\mathrm{\AA}$ line in the second order which is offset from the first order, but presents such a large coincidence loss pattern that it causes a re-distribution of the line flux (e.g. \citealt{Kuin_etal_2015}).  We measured the flux across the pattern. Figure~\ref{Fig:UV_line_flux} illustrates the evolution of the line fluxes with time.

The \eal{Mg}{II} 2800\,$\mathrm{\AA}$ and \hfeal{C}{III} 1909\,$\mathrm{\AA}$ were compared to establish that no coincidence loss affects the \hfeal{C}{III} 1909\,$\mathrm{\AA}$ line fluxes. The reason we believe that coincidence loss is corrected appropriately is that the \eal{Mg}{II} 2800\,$\mathrm{\AA}$ line flux is comparable and lies in the much more sensitive  range of the grism where it would be affected more.

Having established this, the initial rise in the \hfeal{C}{III} 1909\,$\mathrm{\AA}$ flux prior to day 16 is real and different from the flux changes seen in the \eal{Mg}{II} 2800\,$\mathrm{\AA}$ resonance line. Other resonance transitions are present, \eal{N}{II} 2143\,$\mathrm{\AA}$, \eal{C}{II} 2324\,$\mathrm{\AA}$, and \eal{Al}{III} 1860\,$\mathrm{\AA}$, and they all show a similar behaviour, where the line flux shows a decrease from the first spectrum observed at day 6. 

In the \eal{O}{III} 3130\,$\mathrm{\AA}$ line, which is probably optically thin because it is caused by fluorescence from the \eal{He}{II} 304 EUV line, we observe a nearly straight slope, whereas in the permitted lines the flux decrease is initially steeper to join in the same rate of decrease as all lines some time after day 20. As in the photometric data, after day 60 the slope changes, and the line flux evolution is seen to be consistent with the photometric light-curve (see Figure~\ref{Fig:uvot_lightcurve}) taking into consideration the large uncertainties due to the large S/N in the late time spectra. 

The different temporal development of the line fluxes is likely due to opacity. In the resonance lines we can approximate the emission originating from an optical depth of one. Comparison of the temporal development of the optically thick lines to the optically thin line can be  interpreted that the initial flux in the optically thick lines is higher. This can be explained if the flux source of the optically thick lines is dependent on the earlier and higher flux. Assuming that the ejecta are moving at a nearly constant speed, so that there are no frequency shifts, the optically thick line may `bottle-up' some of the earlier high flux radiation. The most likely mechanism is back-scatter of the WD radiation by the ejecta. By that means the radiation density in the line is enhanced within the confines of the ejecta and the flux in the line is partly from the earlier radiation density. When the line becomes effectively optically thin (i.e., photons will escape through the shell, even when the optical depth is still larger than one), the temporal development of the line flux in the optically thick lines will start to match the optically thin line.

What this tells us is that the ejecta up to at least day 20 are not transparent, consistent with the SSS flux not being observed until day 28. It also suggests that the central source is still mostly covered by optically thick ejecta in all directions in these lines, i.e, there is no large hole in the ejecta letting the back-scattered radiation field escape before that time.  In the context of asymmetric ejecta \citep{Shore_etal_2011}, this means that at least some matter is ejected in all directions, though the mass in certain directions may be much larger.

\section{Discussion}
\label{Disc}

\subsection{Distance Modulus}
\label{dist}
While the mean distance to the SMC is around 61 kpc this galaxy is very
extended in the line-of-sight \citep{Caldwell_etal_1986,Jacyszyn_etal_2016}, with
Cepheid variables distributed from $<$54\,kpc to $>$76\,kpc. In this Section we will try to derive an estimate of the distance to the nova using the available methods and test at the same time the usefulness of these methods. 

\begin{figure}
\centering
  \includegraphics[width=\columnwidth]{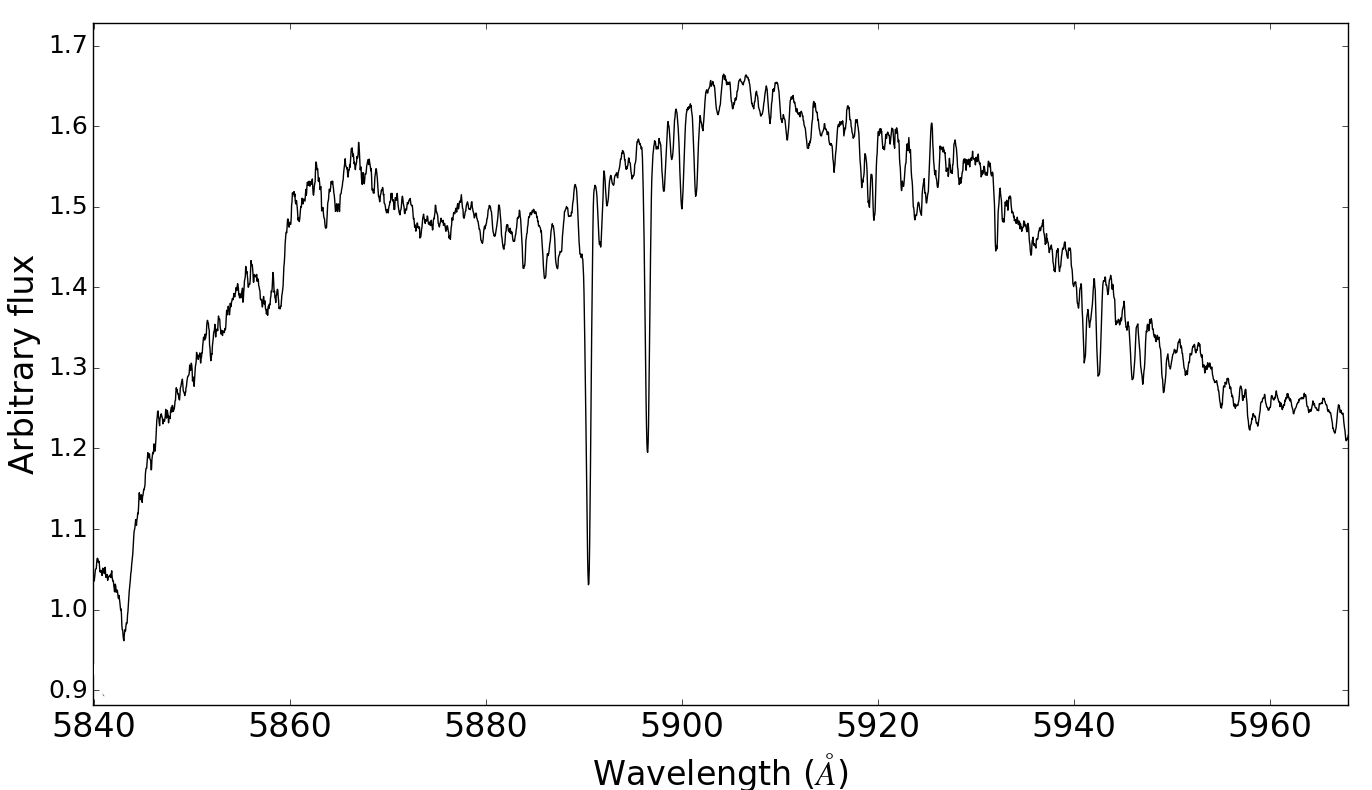}
\caption{The \eal{Na}{I}D doublet absorption lines D1 (5896.6\,$\mathrm{\AA}$) and D2 (5890.6\,$\mathrm{\AA}$) from the high resolution spectrum of day 8.}
\label{Fig:NaD}
\end{figure}

 \citet{Buscombe_1955} suggested that all novae have a similar absolute magnitude 15 days after eruption. \citet{Cohen_1985} estimated this as  
$M_{V,15} = -$5.60\,$\pm$\,0.45 using 11 objects, while
\citet{Downes_etal_2000} refined the value to $M_{V,15} = -6.05 \pm
0.44$ using 28 objects. 
Interstellar reddening towards the SMC is small and for our analysis we assume
$A_V=0.11\pm 0.06$ and $A_I=0.07\pm 0.06$, using relations (4) and (5) in \citet{Haschke_etal_2012} and the average value of $E(V-I)$ for the SMC from \citet{Haschke_etal_2011}. 
Using the EWs of the \eal{Na}{I}D absorption lines (Figure~\ref{Fig:NaD}  - EW$_{\mathrm{D1}}$ = 0.13$\pm$0.02 and EW$_{\mathrm{D2}}$= 0.17$\pm$0.02), and their relationship to the interstellar extinction (as specified by \citealt{Munari_Zwitter_1997} and \citealt{Poznanski_etal_2012}) we obtain values of $A_V$ consistent with our assumptions. Most likely the reddening contribution from the SMC is very low (see Section~\ref{brightness} and the discussion about the radial velocities of the \eal{Na}{I}D absorption lines).

At day 15 from the eruption this SMC nova was at an apparent magnitude $V = 12.30 \pm 0.01$ or $V_0= V - A_V = 12.19\pm0.07$. Using the
\citet{Downes_etal_2000} relation this gives  $(m - M)_0 = 18.23 \pm
0.51$ and a distance $d \sim 44 ^{+11}_{-9}$\,kpc.

After a survey of 24 novae in M87, \citet{Shara_etal_2017_III} recently found that the minimum scatter of novae absolute magnitude occurs at day 17 in the $V$-band and day 20 in the $I$-band. Using the Hubble Space Telescope F606W and F814W filters for the $V$ and $I$ bands respectively, they found $M_{V,17} = -$6.06\,$\pm$\,0.23 and $M_{I,20} = -$6.11\,$\pm$\,0.34. At day 17 from maximum light, nova SMCN 2016-10a was at $V$ = 12.60\,$\pm$\,0.01 or $V_0$ = 12.49\,$\pm$\,0.07. Using the absolute magnitude at day 17 from \citet{Shara_etal_2017_III} gives $(m - M)_0$ = 18.55\,$\pm$\,0.30 and a distance $d \sim 51 ^{+7}_{-6}$\,kpc.
At day 20 from maximum light, nova SMCN 2016-10a was at $I = 12.08 \pm 0.01$ or $I_0= I - A_I = 12.01 \pm 0.07$. Using the absolute magnitude at day 20 from \citet{Shara_etal_2017_III} gives $(m - M)_0 = 18.12 \pm 0.41$ and a distance $d \sim 42 ^{+8}_{-7}$\,kpc.

The nova is located about two degrees south of the central body of the SMC and outside of the region examined by \citet{Subramanian_etal_2017}. Nevertheless, it is worth
noting that these authors identify a red clump population at $\sim$ 12 $\pm$ 2\,kpc in front of the main body of the SMC, that may have been the result of tidal stripping. Other explanations for the discrepancies in the distance: (1) the validity of using $M_{V,15}$ as a distance indicator which was questioned by several authors (see e.g. \citealt{Jacoby_etal_1992,Ferrarese_etal_2003,Darnley_etal_2006}); (2) These relations are supposed to give a rough estimate of the distance and not accurate values; (3) underestimating the absolute magnitude derived using the relations of \citet{Downes_etal_2000} and \citet{Shara_etal_2017_III}. The latter derived their relations using a sample of novae in M87, hence, a parent galaxy with different metallicity than the SMC; (4) Although, the extinction is not expected to play a major role relative to the direction of the nova, an overestimation in the extinction values adopted can lead to a slight underestimation in the distance ($\sim$ 2\,kpc). We conclude that the nova is within the SMC and possibly foreground of the bulk of the stars in the field.  A less likely possibility that the nova belongs to the Galaxy is discussed in Section~\ref{brightness}. On the assumption that it is associated with the SMC we will adopt a distance of 61\,$\pm$\,10\,kpc throughout the rest of the paper. This distance covers the extended SMC and the distances we derived under uncertainties.

\subsection{The progenitor system}

\begin{figure}
\begin{center}
  \includegraphics[width=\columnwidth]{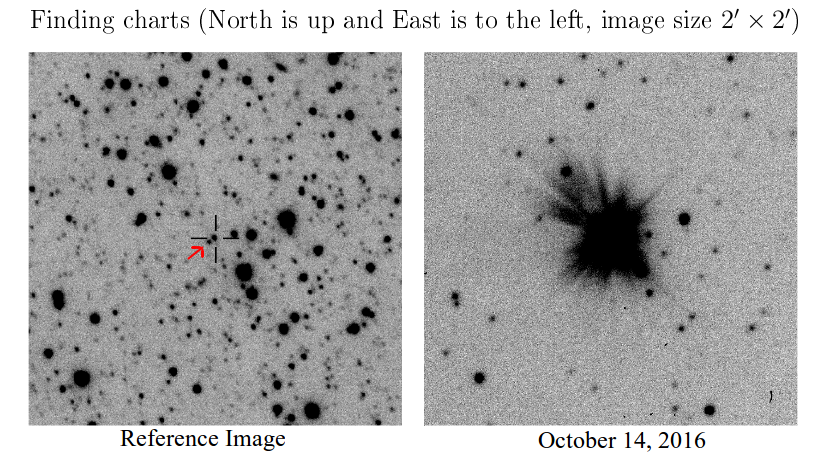}
\caption{Left: OGLE finding chart of the progenitor system before the eruption. Right: image of the field 5 days after the eruption.}
\label{Fig:finding_chart}
\end{center}
\end{figure}

During quiescence, the emission from a CN system is mainly from the secondary star and the accretion disk (depending on its inclination relative to the line of sight). The emission from the WD is almost negligible during quiescence. The OGLE survey has been monitoring the target since 2010. Figure~\ref{Fig:finding_chart} shows a finding chart of the object during quiescence (left) and an image of the object during the eruption (right). The star South-East of the nova (indicated with the red arrow) is $\sim$ 2.1 arcsec away. 

\begin{figure}
\begin{center}
  \includegraphics[width=\columnwidth]{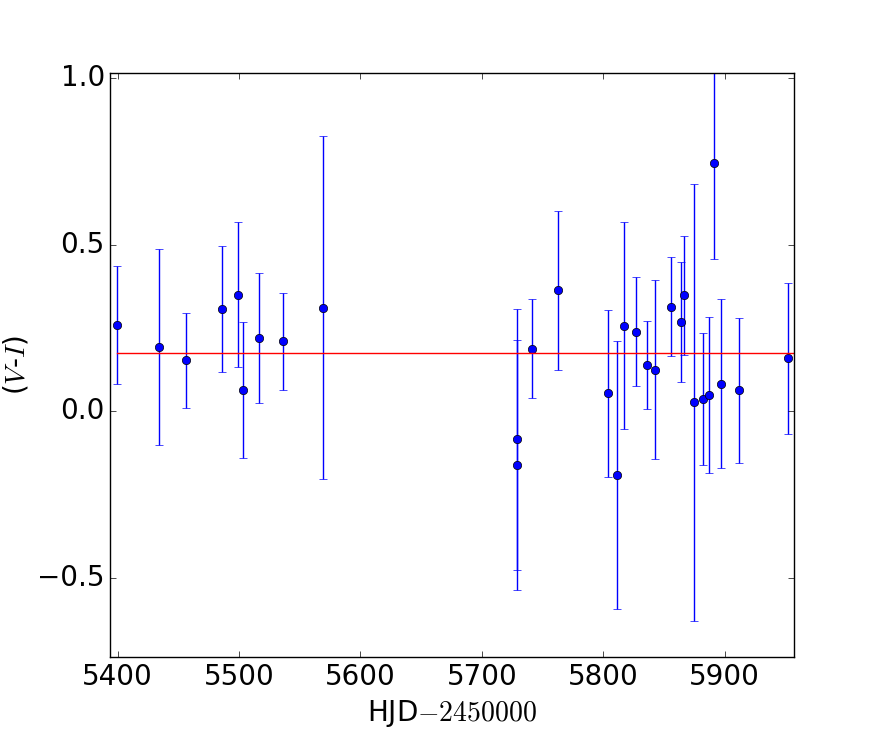}
\caption{The evolution of $(V-I)$ during quiescence. The red line presents the mean values of $(V-I)$.}
\label{Fig:v-i}
\end{center}
\end{figure}

In quiescence, the target shows irregular variability that might be due to variations in the mass-transfer rate (see Figure~\ref{Fig:OGLE_LC}). No periodic variability appears in the light-curve of the progenitor that can lead us to constrain the orbital period of the system. This means that the orbital plane of the system may be oriented face-on to the observer, so that the contribution of the disk to the total emission from the system might be large. During quiescence, the progenitor is observed at $\stretchleftright{\langle}{V}{\rangle} \sim 20.70$ and $\stretchleftright{\langle}{I}{\rangle} \sim 20.55$. In Figure~\ref{Fig:v-i} we present the variation of $V-I$ over time during quiescence. We only include the days when the magnitudes of both broadband colours are measured quasi-simultaneously. The $V-I$ colours are more sensitive to the disk than to the secondary, if the latter is a main sequence or sub-giant. However, these colours would be equally sensitive to a red giant secondary. The changes in $V$ and $I$ during quiescence (Figure~\ref{Fig:OGLE_LC}) and the moderately high values of $V-I$ might reflect the contribution of a bright disk that is showing temporal variability.\\

\begin{figure*}
\centering
\begin{subfigure}{0.45\textwidth}
\centering
\includegraphics[width = \textwidth]{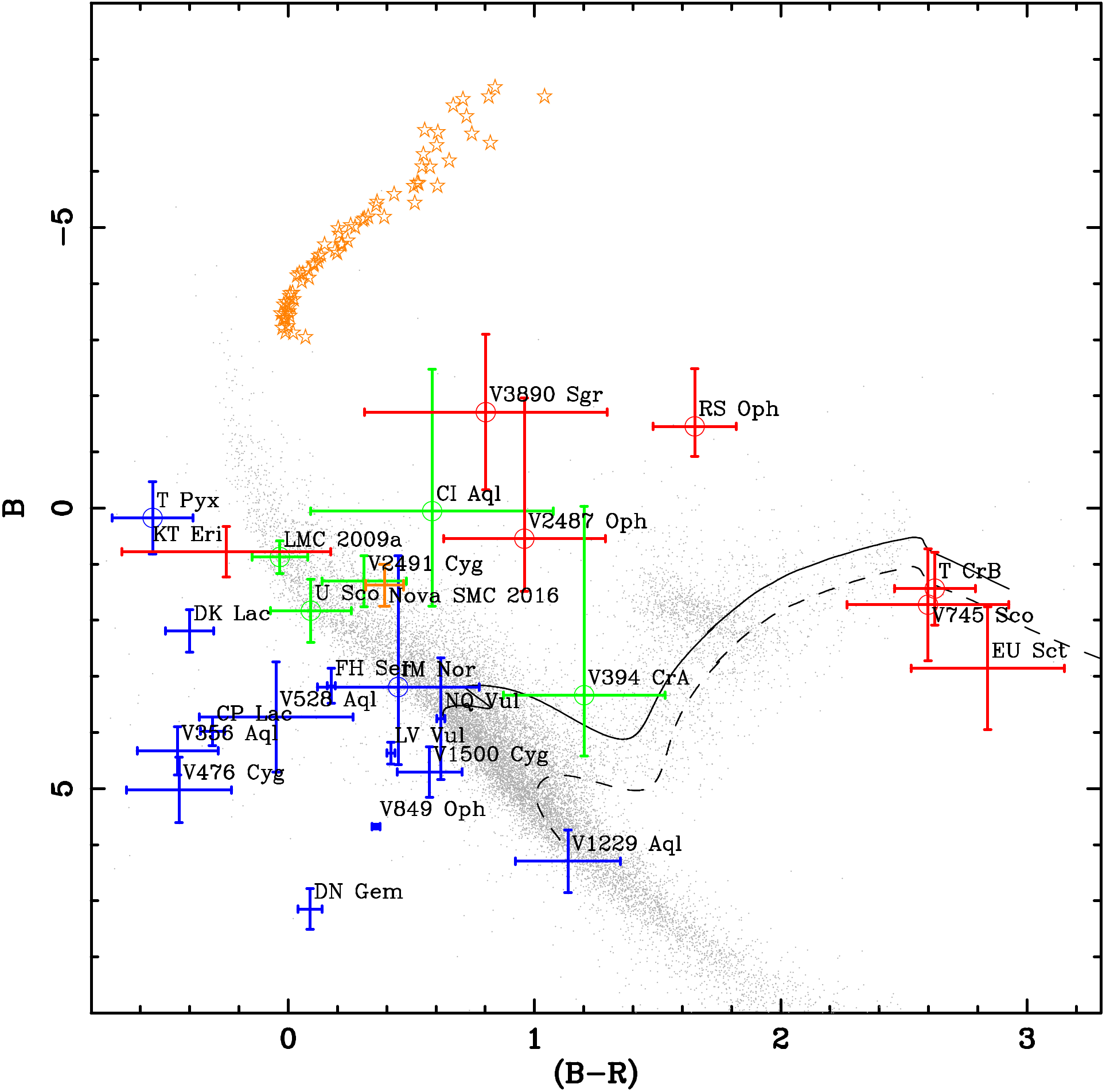}
\label{fig:left}
\end{subfigure}
\begin{subfigure}{0.45\textwidth}
\centering
\includegraphics[width = \textwidth]{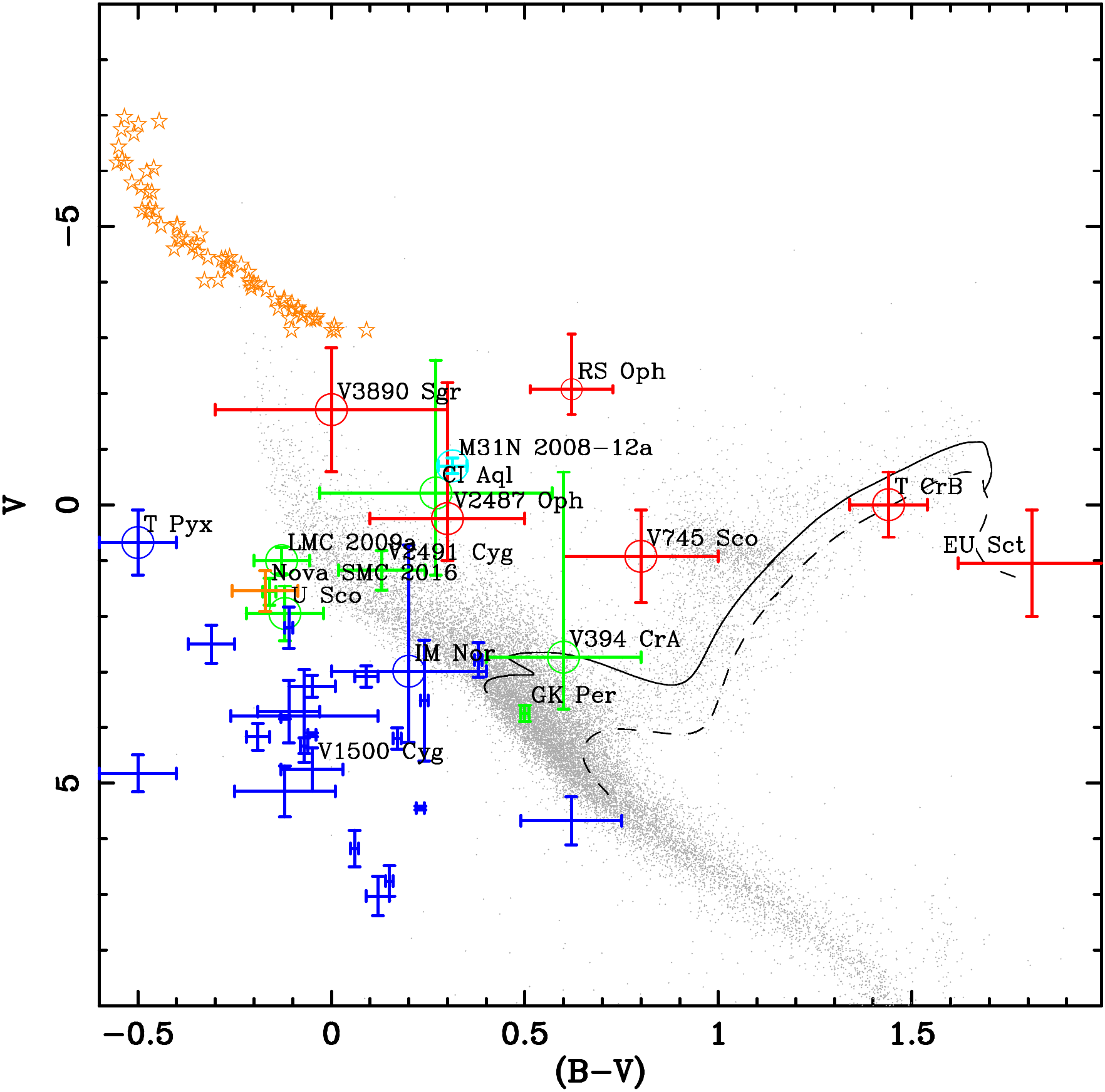}
\label{fig:right}
\end{subfigure}
\begin{subfigure}{0.45\textwidth}
\centering
\includegraphics[width = \textwidth]{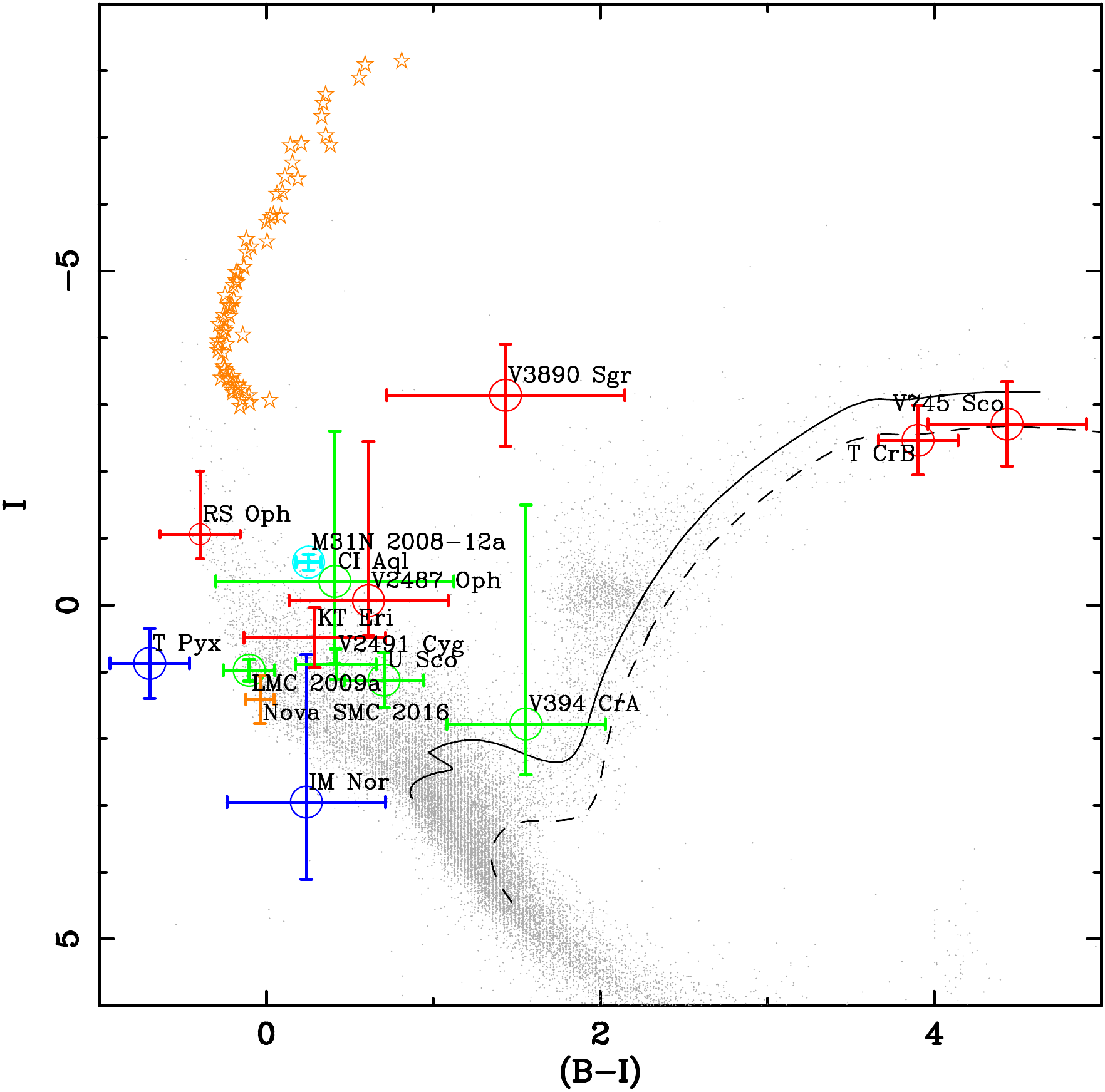}
\label{fig:left}
\end{subfigure}
\begin{subfigure}{0.45\textwidth}
\centering
\includegraphics[width = \textwidth]{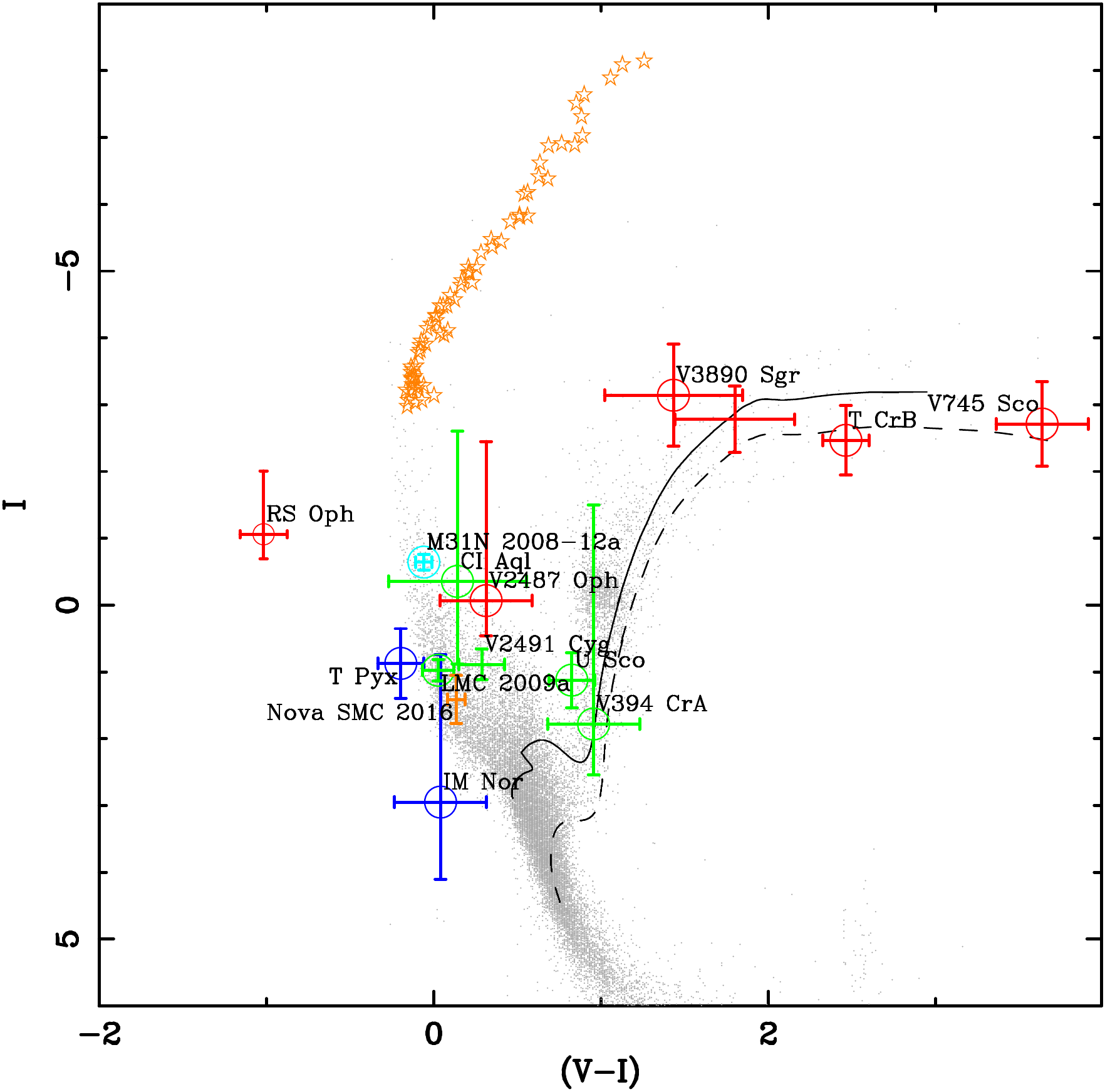}
\label{fig:right}
\end{subfigure}
\caption{Colour-magnitude diagrams showing stars (grey dots) from \textit{Hipparcos} data set \citep{Perryman_1997} with parallax errors $<$\,10\%. These stars have been transformed to the distance and extinction of the SMC. \textit{BVI} photometry is taken directly from the \textit{Hipparcos} catalogue, $R$ photometry is taken from the USNO-B1 catalogue (\citealt{Monet_etal_2003}; via the VizieR database, \citealt{Ochsenbein_etal_2000}). The blue points represent Galactic MS-novae, the green points represent Galactic SG-novae and red points represent Galactic red giant novae (RG-novae) (see \citealt{Schaefer_2010,Darnley_etal_2012}, and references therein). The known recurrent novae  in this sample have been identified with an additional circle. Orange points represent nova SMCN 2016-10a during quiescence. The orange stars indicates the track of nova SMCN 2016-10a from day 11 till day 108 post-eruption. M31N 2008-12a is shown in light blue due to its uncertain donor. A colour version of this plot is present in the online journal.} 
\label{CMDs}
\end{figure*}

We found several sources less than $\sim$\,1.0${\arcsec}$ from the nova in archival data from: USNO-B1 catalogue \citep{Monet_etal_2003} ($B = 20.17$ and $R = 20.0$); the Magellanic Cloud Photometric Survey (MCPS; \citealt{Zaritsky_etal_2002}) ($U$ = 19.71\,$\pm$0.14, $B$ = 20.44\,$\pm$0.07, and  $V$ = 20.58\,$\pm$0.04); GALEX catalogues \citep{Bianchi_etal_2011} (NUV = 20.57\,$\pm$\,0.20 and FUV = 20.53$\pm$0.33). After checking several charts and images of the field from OGLE, MASTER and Simbad \citep{Wenger_etal_2000}, it seems likely that all these observations refer to the nova progenitor. These data along with the OGLE data, allow us to place the progenitor on $B$ versus $B-R$, $V$ versus $B-V$, $I$ versus $B-I$ and $I$ versus $V-I$  colour-magnitude diagrams (CMDs, Figure~\ref{CMDs}). The $B$ versus $B-R$ indicates a system which may have a sub-giant companion (SG-novae). However, the three other CMDs indicate a system with a main sequence companion (MS-novae). The inconsistency of the $B$ versus $B-R$ with the other plots could be understood if the progenitor had strong H$\alpha$ emission.
It is worth noting that $B$, $V$, $R$, and $I$ bands are very sensitive to the emission from the disk. The optical colours depend sensitively on the ratio of the disk to stellar photospheric emission.

\citet{Darnley_etal_2012} show that, ideally, placing the system on a CMD using NIR colours can give us an insight into the secondary star type. The NIR colours are more sensitive to the emission from the secondary than that from the disk. With lack of observations in NIR for the progenitor in the archival data, it is difficult to say much about the secondary star.

\begin{figure*}
\begin{center}
  \includegraphics[width=\textwidth]{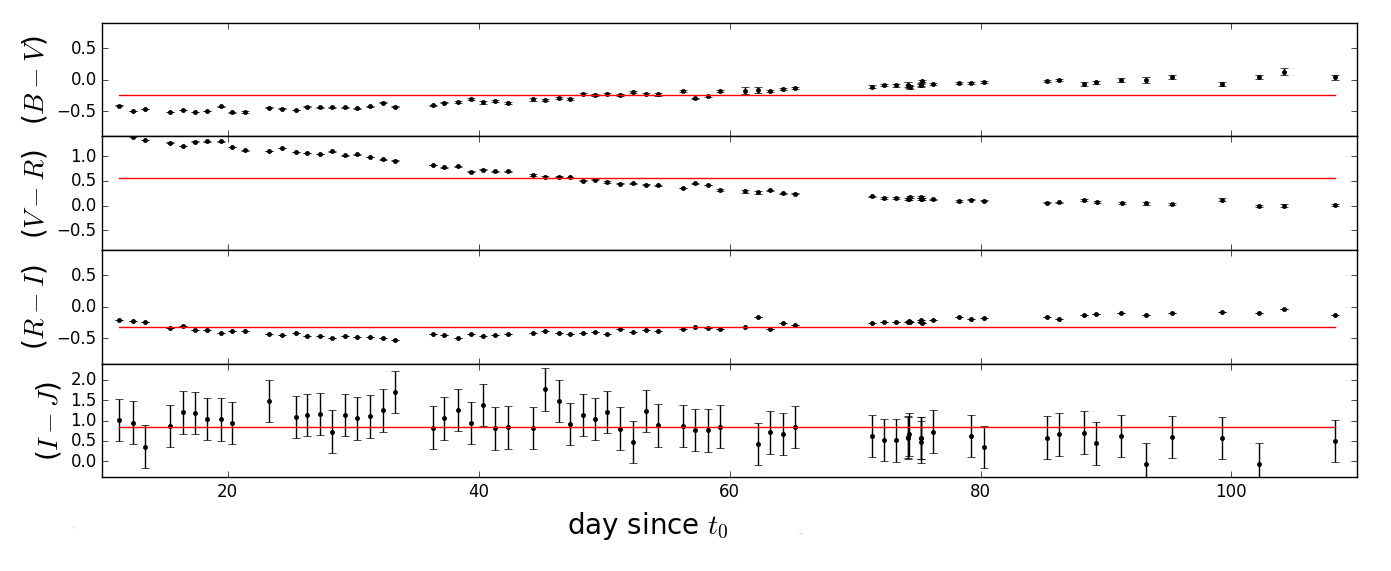}
\caption{Optical and NIR broadband colour evolution of nova SMCN 2016-10a between day $\sim$11 and day $\sim$100 post-eruption. The red line is the average of the values between the two dates.}
\label{Fig:color_evolution}
\end{center}
\end{figure*}

\subsection{Eruption and early decline}

The system was last observed by OGLE before the eruption on 2457667.67 at $V = 20.65 \pm 0.03$. Adopting a distance of $61\pm10$\,kpc and $A_V = 0.11\pm 0.06$ will result in $M_V = + 1.61 \pm 0.45$ three days before the eruption. Although the MASTER VWF camera measurements are not precise $V$-band photometry, the measurements can be considered as an approximation of $V$-band photometry (see Section~\ref{LC}). Hence, we consider $V_{\mathrm{max}}\approx 8.5$\,$\pm$\,0.1 which results in $M_{V,\mathrm{max}}$\,$\approx$\,$-10.5$\,$\pm$\,0.5. We can thus constrain an approximation for the eruption amplitude as $A$\,$\approx$\,12.1\,$\pm$\,1.0. 
The amplitude of the eruption is not only related to the eruption itself, but also to the luminosity of the donor during quiescence. Hence, such a high eruption amplitude might also indicate a sub-luminous donor, such as a main sequence one. The eruption amplitude agrees well with the amplitudes given in the $A$ versus $t_2$ relationship for CNe in \citet{Warner_Brian_1995}, where CNe of similar speed class to SMCN 2016-10a, with a main sequence donor, are characterized by $A$\,$\geq$\,12. 

\subsubsection{The brightest nova discovered in the SMC?}
\label{brightness}

If nova SMCN 2016-10a is indeed in the SMC, $M_{V,\mathrm{max}}$\,$\approx$\,$-10.5$\,$\pm$\,0.5 will make it the brightest nova discovered in the SMC, and probably one of the brightest ever discovered. At $M_{V,\mathrm{max}}$\,$\approx$\,$-10.5$\,$\pm$\,0.5 it is at least as bright as V1500 Cyg and CP Pup. These two novae are considered the brightest nova events ever recorded (see e.g. \citealt{Warner_1985,Stockman_etal_1988,Hachisu_etal_2006,Shafter_etal_2009}). If the nova is in the foreground of the SMC at $\sim$\,46\,$\pm$\,10\,kpc (the mean value of the three distances derived in Section~\ref{dist}), we derive $M_{V\mathrm{max}}$\,$\approx$\,$-$9.9\,$\pm$\,0.6 meaning it is still a very luminous nova ($M_{V\mathrm{max}} \leq -$9.0; \citealt{Shafter_etal_2009}). If the nova belongs to our Galaxy, which is very much less likely, then $M_{V\mathrm{max}}$ is greater than $-$6.0. In this case the nova may be under-luminous and possibly classified as a faint fast nova (see e.g. \citealt{Kasliwal_etal_2011,Shara_etal_2017_II}).

The first value is in good agreement with the SMARTS $V$-band photometry. The nova magnitude in the $V$-band has been measured as 12.14 ($V_0$ = 12.03\,$\pm$\,0.06) at $t=11.3$\,d. We estimate $t_3 = 7.8 \pm 2.0$\,d  which is $<$ 11.3\,d, meaning that the nova reached a maximum brightness in the $V$-band smaller than 12.03\,$\pm$\,0.06 $-$ 3.0 = 9.03\,$\pm$\,0.06. This value leads us to constrain a maximum on $M_{V,\mathrm{max}}$\,$<$\,$-9.9$\,$\pm$\,0.5 (at $d$ = $61\pm10$\,kpc). 

In order to constrain the distance to the nova, we derived radial velocities from the \eal{Na}{I}\,D doublet absorption lines at 5890\,$\mathrm{\AA}$ and 5896\,$\mathrm{\AA}$. The lines are detected at $\sim$ 5890.6\,$\mathrm{\AA}$ and $\sim$ 5896.6\,$\mathrm{\AA}$, showing a red-shift of $\sim$\,16\,$\pm$\,2\,km\,s$^{-1}$ in all five HRS spectra. The origin of these lines is either circumstellar or interstellar material. The line widths are too narrow to be produced in an outflow and hence they come from a static region (see Figure~\ref{Fig:NaD}). Since the average radial velocity of the SMC is $\sim$\,160\,km\,s$^{-1}$ \citep{Neugent_etal_2010}, these absorption lines are due to Galactic interstellar medium absorption. Note that if these absorption lines originate from the SMC, they should be at $\sim$ 5893\,$\mathrm{\AA}$ and $\sim$ 5899\,$\mathrm{\AA}$. In conclusion, none of the three possibilities of the distance to the nova can be ruled out. But it is still very likely that the  nova is in the SMC or foreground of the SMC, and the interstellar absorption is from the Milky Way. 

\subsubsection{Post-eruption colour evolution}
In Figure~\ref{Fig:color_evolution} we present the evolution of  different broadband colours during the early decline until $\sim$ 100 days post-eruption. We only include the days when the magnitudes of different broadband colours are measured quasi-simultaneously. 

For objects that show emission lines in their spectra, the emission lines can be a significant contributor to the flux. Hence, the evolution of the colours is an indication of the evolution of both the lines and the continuum \citep{Bode_etal_2016}. In the optical, the emission lines contribute strongly to the $B$-band flux, especially the Balmer lines (H${\beta}$, H${\gamma}$,  H${\delta}$, and H${\epsilon}$) during the early days after the eruption. In the $R$-band, H${\alpha}$ is a strong contributor to the flux. The \eal{O}{I} lines are also an important contributor to the $I$-band flux during the early decline. The $B-V$ broadband colour evolution starts bluer at $\sim$ 11 days after eruption, showing a monotonic evolution toward redder colours until 100 days post-eruption. This evolution of $B-V$ can be explained by the dimming of the Balmer emission lines during the decline from maximum light which affect the flux in the $B$-band, followed by the emergence of the nebular \feal{O}{III} lines, at the blue edge of the $V$ band, after the start of the nebular phase (see section~\ref{spec_class}). 

The evolution of $V-R$ starts redder and becomes bluer monotonically with time, which in turn can also be attributed to the dimming of H${\alpha}$. The $R-I$ evolution shows slight variation around the mean, probably due to the relative variation of the \eal{O}{I} lines compared to H${\alpha}$. In the NIR, the evolution of the broadband colours  show no trend with time. They show slight random changes around the mean, but the uncertainty on the data is very high and hence it is difficult to say anything about the variation of these colours.

\citet{Hachisu_etal_2016} have demonstrated that the evolution of the colours, during and after the eruption, is different for systems hosting a red giant companion (RG-novae) than for SG-novae or MS-novae. They showed that for the RG-novae, the colour evolution follows a vertical track, on the CMD, along the two lines of optically thick wind and optically thin wind free-free emission intrinsic colours, respectively $(B - V)_0 = -$0.03 and $(B-V)_0$ = 0.13 (see also \citealt{Hachisu_etal_2014,Darnley_etal_2016}). For SG- and MS-novae, the colour-magnitude tracks show an evolution towards bluer colour to reach a turning point, close to the start of the nebular phase, from where the colours evolve backward and become redder.

\begin{figure}
\centering
  \includegraphics[width=\columnwidth]{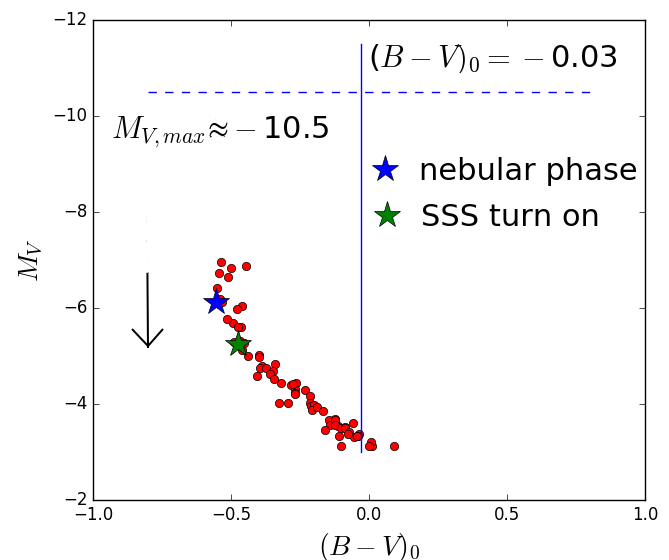}
\caption{Colour-magnitude diagram of SMCN 2016-10a showing $M_V$ against $(B-V)_0$ (red dots) at a distances $d = 61$\,kpc. The evolution spans from day 12 to day $\sim$100 post-eruption. The black arrow indicates the evolution of $M_V$ with time.  The blue horizontal dashed line indicates the maximum absolute magnitude. The blue vertical line indicates the intrinsic colours for an optically thick wind free-free emission (see \citealt{Hachisu_etal_2014} and \citealt{Darnley_etal_2016} for further details). The green and blue stars represent the start of the SSS phase (day 28) and the nebular phase (day 20) respectively. A colour version of this plot is present in the online journal.}
\label{Fig:colomagdia}
\end{figure}

In Figure~\ref{Fig:colomagdia} we present a CMD, illustrating the evolution of $M_V$ against $(B-V)_0$, for nova SMCN 2016-10a. The track shows an evolution similar to novae with a sub-giant or a main sequence secondary (see figure 7 in \citealt{Darnley_etal_2016}). Although we lack broadband observations during the first 10 days after maximum optical light, the available data show a blue-ward evolution before the turning point into a red-ward evolution near the start of the nebular phase. The evolution of the CMD can be interpreted as follows: at the very early phase after maximum optical peak, the colours show a gradual change towards the blue, corresponding to the free-free emission phase of the nova spectrum. Later, after maximum, the colours becomes bluer due to the strong emission lines within the $B$-band. Then, these lines start to fade and the turning point starts. Close to the start of the nebular phase, the colours turn red-ward due to the emergence of nebular lines such as \feal{O}{III} 4959\,$\mathrm{\AA}$ and 5007\,$\mathrm{\AA}$ (see Section~\ref{spec_class}) at the blue edge of the $V$-band \citep{Hachisu_etal_2016}.

\begin{figure*}
\centering
  \includegraphics[width=\textwidth]{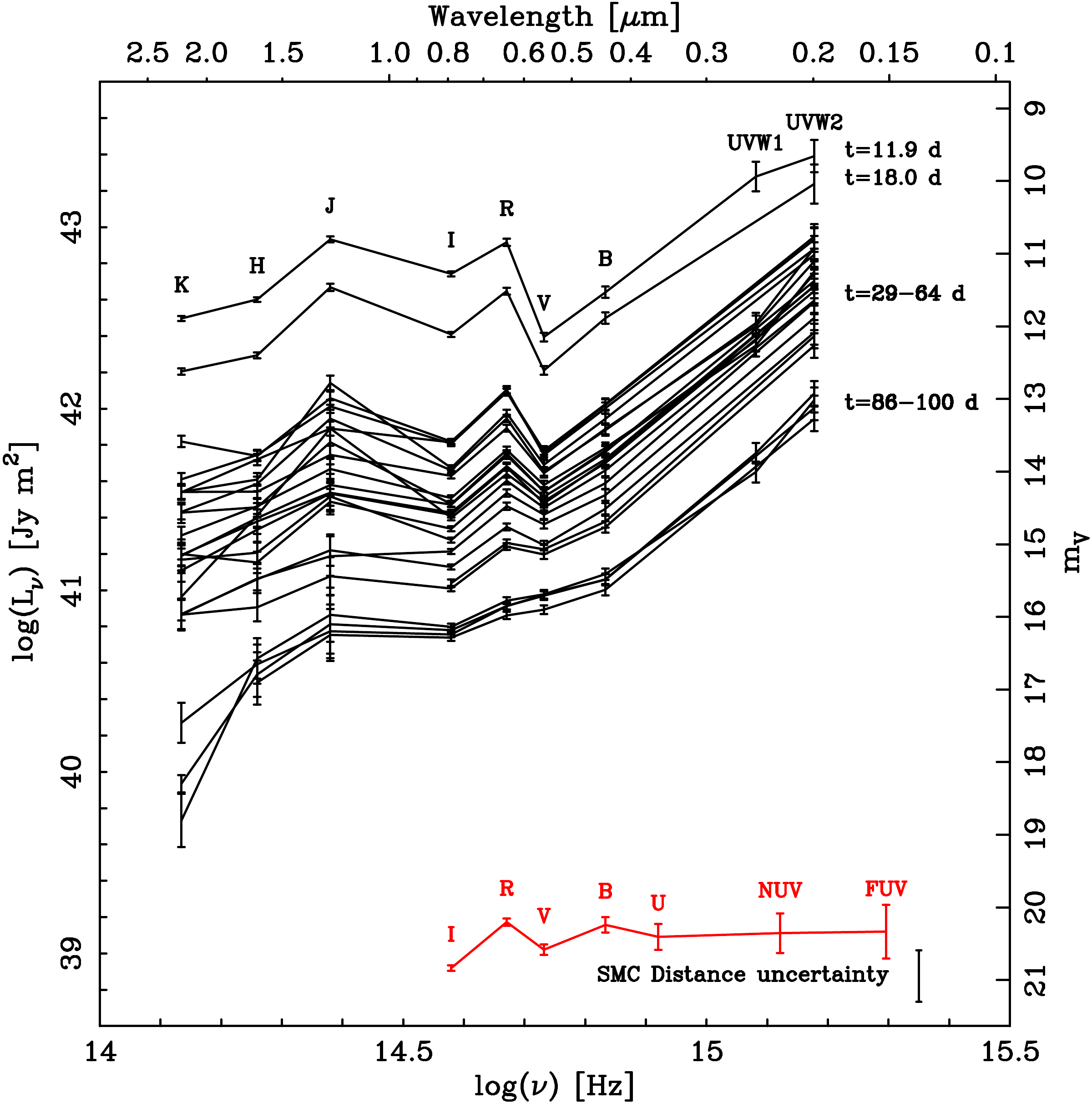}
\caption{Distance- and extinction-corrected SED plots showing the evolution
of the SED of nova SMCN 2016-10a eruption. The error bars include contributions
from the photometric and distance uncertainties. Red data at the bottom present the quiescence SED, using the MCPS $UBV$, USNO-B1 $R$, OGLE $I$ and GALEX NUV/FUV.}
\label{Fig:SED}
\end{figure*}

Figure~\ref{Fig:SED} illustrates distance- and extinction- corrected spectral energy distribution (SED) plots, showing the evolution of the post-eruption SED of nova SMCN-2016-10a. SMARTS and {\em Swift} UVOT data are combined to form the SED across the NIR, optical, and UV spectral range. We only plot the data that have corresponding observations in all the bands. In the optical, the evolution of the SED plots is affected by the line emission as discussed for the CMD and colour evolution plots. H$\alpha$ contributes strongly in the, initially bright, $R$ band. Since nova behaviour during and after peak is similar to blackbody emission (see e.g. \citealt{Gallagher_Ney_1976,Bode_etal_2008}), we can conclude from Figure~\ref{Fig:SED} that the peak of emission is beyond the uvw2 of {\em Swift} UVOT (shorter than 2000\,$\mathrm{\AA}$) throughout the 12-100 day interval. At optical maximum the spectra of CNe show a blackbody emission peaking in the optical with temperatures around 10,000\,K (see e.g. \citealt{Bode_etal_2008} and references therein). A simplified model of an evolving pseudo-photosphere, in the optically thick ejecta, has been adopted to re-produce the evolution of the post-eruption emission from novae  (see e.g. \citealt{Bath_Harkness_1989}). The temperature of the photosphere reaches its lowest at the optical peak, and then increases with time while the radius of the photosphere shrinks. This increase in the temperature shifts the blackbody temperature towards the UV. Since nova SMCN 2016-10a is a very fast nova and since the first SED plot is around 12 days post-optical-maximum, it is expected that the SED peaks toward the UV. This peak is beyond the uvw2 of the {\em Swift} UVOT, which indicates temperatures above 15,000\,K.

\subsection{Mass of the WD and the ejected envelope}
\label{MWD}
It is now well understood that several parameters other than the WD mass affect nova eruptions, including the accretion rate on the WD surface, the WD temperature (and thus luminosity), the WD chemical composition, and the chemical composition of the accreted matter (see e.g. \citealt{Yaron_etal_2005,Hillman_etal_2014,Shara_etal_2017_II}, and references therein). However, it is widely accepted that the WD mass is the critical factor in nova eruptions, and it has a great effect on the light-curve morphology and decline rate. Novae with high mass WDs ($M_{\mathrm{WD}} >$ 1.0\,M$_{\odot}$) are expected to show fast declines in the optical light, while novae with low mass WDs are expected to show slow optical decline rates. The \citet{Hillman_etal_2016} models predicted that any novae with $t_2 <$ 10 days must contain a WD with a mass $\ge$ 1.25 M$_{\odot}$.

The mass of the accreted envelope is also related to the decline rate and the WD mass.
We use the empirical relation from \citet{Shara_etal_2017_II} to estimate the accreted envelope mass: $$\log M_{\mathrm{env}} = 0.825 \log(t_2) - 6.18\,,$$
resulting in $M_{\mathrm{env}} = (0.20 \pm 0.05) \times 10^{-5}$\,M$_{\odot}$, which is in good agreement with \citet{Hachisu_etal_2016} models for fast novae. One should note that the empirical relation presented in \citet{Shara_etal_2017_II} was derived by fitting novae from M87, M31, LMC, and the Milky Way.

\subsubsection{SSS and $M_{\mathrm{WD}}$}
The timescale of the SSS phase and uncovering the X-ray source is proportional to the hydrogen ejected envelope mass and inversely proportional to the ejection velocity. \citet{Krautter_etal_1996} relate the SSS phase turn-on to the mass of the ejected H envelope and the ejecta velocity by: $t_{\mathrm{on}} \propto M_{\mathrm{H}}^{1/2} v_{\mathrm{ej}}^{-1}$. With relatively low envelope mass and high ejecta velocity, the time since eruption to observe the SSS phase should be relatively short, which is the case for nova SMCN 2016-10a with $t_{\mathrm{on}} =$ 28\,days. This value is comparable to other novae of the same speed class (see e.g. \citealt{Bode_etal_2016} and references therein). $t_{\mathrm{on}}$ is dependent on the mass of the WD, hence for systems with high M$_{\mathrm{WD}}$, $t_{\mathrm{on}}$ is expected to be shorter than for  systems with low M$_{\mathrm{WD}}$. The SSS phase of RS Oph \citep{Osborne_etal_2011}, the first nova monitored in detail by {\it Swift}, was first detected on day 26 after eruption, very similar to nova SMCN 2016-10a, while V745 Sco \citep{Page_etal_2015} provided an extreme example, where t$_{on}$ was only 4 d. An extra-galactic perspective is given by nova LMC 2012 \citep{Schwarz_etal_2015}, with t$_{\rm on}$ = 13~$\pm$~5 d. All of these novae are likely examples of high WD mass systems. In comparison, nova SMC 2012 (OGLE-2012-NOVA-002) was a much slower system, not seen as a supersoft source until around nine months after the first optical detection \citep{ATel_4853}.

The turn-off time, since eruption, for nuclear burning ($t_{\mathrm{off}}$) is also strongly dependent on M$_{\mathrm{WD}}$, more precisely it is inversely proportional to the WD mass. \citet{MacDonald_1996} found $t_{\mathrm{off}}$\,$\propto$\,M$_{\mathrm{WD}}^{-6.3}$. Novae with similar $t_{\mathrm{off}}$ to that of the nova SMCN 2016-10a are expected to have a WD mass between 1.1\,M$_{\odot}$ and 1.3\,M$_{\odot}$ (see e.g. \citealt{Krautter_etal_1996,Balman_etal_1998,Osborne_etal_2011} and references therein). Despite RS Oph having a similar t$_{\rm on}$ to SMCN 2016-10a, its t$_{\rm off}$ is noticeably shorter, at $\sim$~58 d \citep{Osborne_etal_2011}, implying the WD in the RS Oph system is more massive (see also \citealt{Brandi_etal_2009,Booth_etal_2016}).

 \citet{Henze_etal_2014_Mar} monitored novae in M31 presenting a series of correlations relating SSS parameters. The values of t$_{\mathrm{on}}$ = 28\,d and t$_{\mathrm{off}}$ = 180\,d lie within the scatter of their results. The luminosity-temperature models in \citet{Wolf_etal_2013} (see also \citealt{Sala_Hernanz_2005}) suggest a high ($\sim$~1.25--1.3 M$_{\odot}$) WD mass for the $\sim$ 90\,eV SSS temperatures measured for nova SMCN 2016-10a. Thus, the X-ray data agree with the conclusion  from the optical that the WD is likely of high mass.

\subsection{Spectroscopic classification}
\label{spec_class}
The spectroscopic and photometric data leave no doubt that the object is a CN eruption.
The spectral evolution of nova SMCN 2016-10a shows initially a He/N spectrum dominated by broad prominent Balmer, He, N, and O emission lines \citep{ATel_9628}. A FWHM of $\sim$ 3500\,km s$^{-1}$ was measured for H${\alpha}$ and H${\beta}$ suggesting expansion velocity of around 1800\,km s$^{-1}$. This moderately high expansion velocity, as well as the broad, jagged profiles of the Balmer emission lines (see section~\ref{Balmer}), are features of the He/N class that is characterized by a HWZI larger than 2500 km s$^{-1}$ \citep{Williams_1992}. The large HWZI values for the He/N novae are due to high ejecta velocities, these arise from the initial explosion on the WD. Before the nebular phase of nova SMCN 2016-10a, we note similarity with the spectrum of KT Eridani, also a very fast He/N nova. However, when KT Eridani enters the nebular phase the blended lines of \eal{N}{III} 4638 and \eal{He}{II} 4686 split and a narrow \eal{He}{II} emerges and grows becoming stronger than H$\beta$ (see figure 2 in \citealt{Imamura_Tanabe_2012}) which is not the case for nova SMCN 2016-10a.

In the spectra of day 20 to 23, the forbidden nebular \feal{O}{III} transitions at 4363\,$\mathrm{\AA}$, 4959\,$\mathrm{\AA}$ and 5007\,$\mathrm{\AA}$ emerge while the nitrogen lines start to fade. The blending of the \feal{O}{III} 5007\,$\mathrm{\AA}$ with the \eal{N}{II} 5001\,$\mathrm{\AA}$ gives an artificially high \feal{O}{III} 5007/4959 flux ratio at this stage. This flux ratio and the present emission lines indicate that the nova is in transition and had started to enter the optically thin nebular phase by the end of October leading to a correct prediction of the imminent SSS turn-on \citep{ATel_9688}. These forbidden lines are expected to initially emerge from low density gas due to the expansion of the nova \citep{Williams_1992}.

\begin{figure*}
  \includegraphics[width=\textwidth]{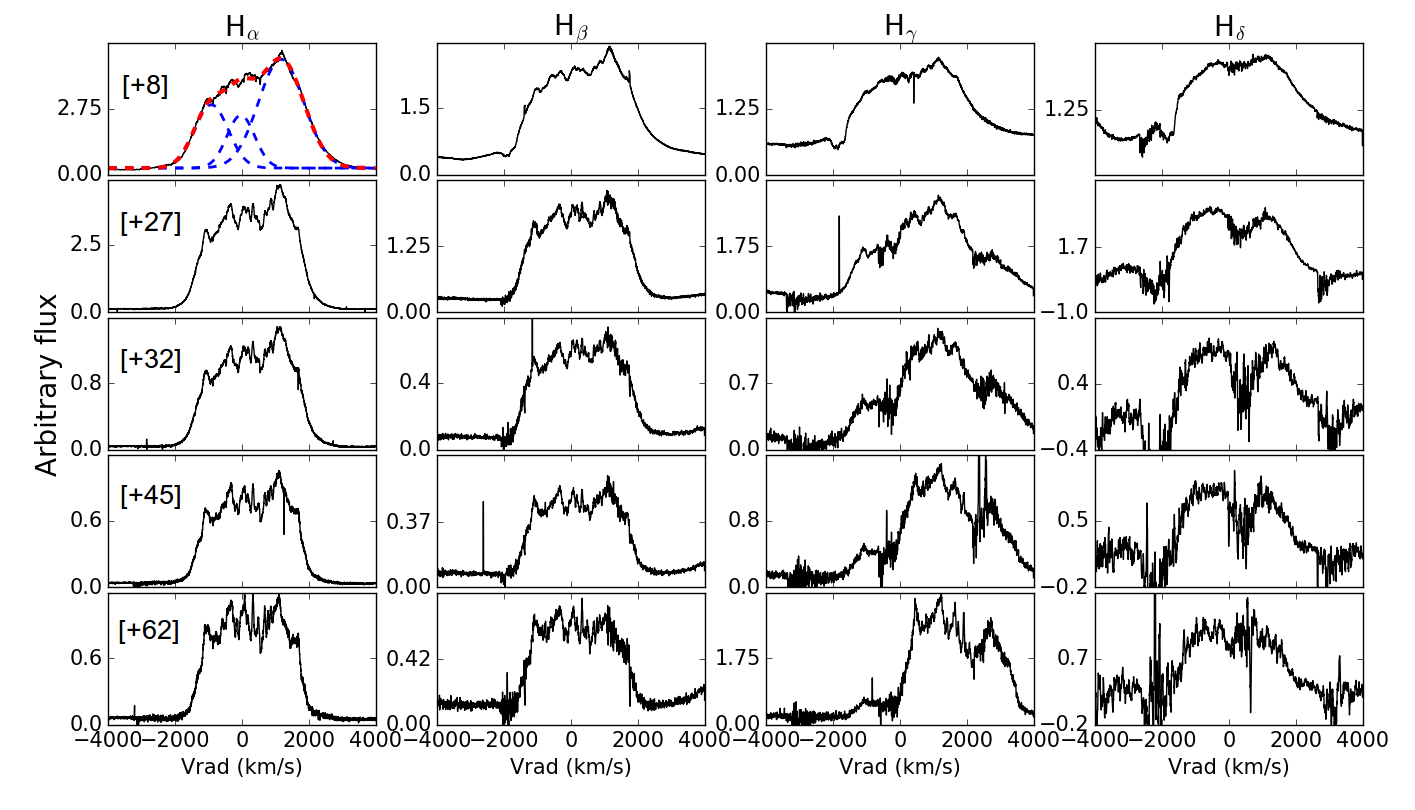}
\caption{The evolution of Balmer lines from SALT high resolution spectra. From left to right: H${\alpha}$, H${\beta}$, H${\gamma}$, and H${\delta}$. The numbers between brackets are days after $t_0$. Heliocentric correction is applied to the radial velocities. The top-left panel shows a sample of multiple Gaussian fitting to the H$\alpha$ line used to derive the FWHM.}
\label{Fig:Balmerlines}
\end{figure*}

The spectra of day 27 and 32 show that the nova had entered the nebular phase with the \eal{He}{II} 4686\,$\mathrm{\AA}$ line becoming stronger than H${\beta}$. According to \citet{Williams_1992}, He/N novae are expected to evolve in three distinct ways in the nebular stage: either showing no forbidden lines while the emission spectra fade into the continuum; coronal lines emerge (e.g. \feal{Fe}{X} 6375\,$\mathrm{\AA}$); or the third scenario, where strong forbidden Ne lines emerge resulting in a Ne nova. The spectra of day 32 (Figure~\ref{Fig:fts2}) show relatively strong \feal{Ne}{V} lines compared to the Balmer and \feal{O}{III}, where the strength of \feal{Ne}{V} 3426\,$\mathrm{\AA}$ is comparable to \feal{O}{III} 5007\,$\mathrm{\AA}$. Some novae show very strong Ne emission, and these novae are considered Ne novae. For these novae, the Ne abundance exceeds solar abundance. Without a detailed calculation of element abundances, it is difficult to determine if the progenitor of nova SMCN 2016-10 contains a Ne WD. According to stellar evolution theory, Ne novae are expected to originate on a high mass WD (close to Chandrasekhar mass, $M_{\mathrm{WD}} \geq$ 1.08 M$_{\odot}$; see e. g. \citealt{Nomoto_1984,Truran_etal_1986,Hurley_etal_2000} and references therein). It is worth noting that \citet{Shara_Prialnik_1994}, showed that it is possible to form a moderate-mass Ne--Mg rich WD, due to a high accretion rate of hydrogen onto the surface of a C--O WD prior to the nova eruption, resulting in a massive Ne-Mg rich layer on the surface of the WD (see also \citealt{Hachisu_Kato_2016_Jan}). It is worth noting that the evolution of H$\gamma$ is strongly affected by the \feal{O}{III} 4363\,$\mathrm{\AA}$ line from day 27 and that the evolution of H$\delta$ is affected by instrumental noise. The late spectra of day 275 shows \feal{O}{III} lines and H$\alpha$ with the absence of forbidden Ne lines.

\subsection{Line profile evolution}
\subsubsection{Balmer lines}
\label{Balmer}
The Balmer lines in the SALT high resolution spectra showed asymmetric profiles soon after maximum light (Figure~\ref{Fig:Balmerlines}).
On day 8, all the Balmer lines showed a red-shifted - relative to the rest wavelength - single-emission peak at around +1200 km s$^{-1}$. This single emission peak might be due to an inclined (relative to the observer) asymmetric ejecta. Aspherical ejecta might also explain the difference between the positive (red) and negative (blue) sides of the line profiles. The relative strength of this emission feature decreases passing from H${\alpha}$ to H${\delta}$. Faint absorption features are observed for H${\beta}$,  H${\gamma}$, and  H${\delta}$. Such absorption features were also observed in the same lines of nova V339 Delphini, also a fast nova \citep{Burlak_etal_2015} that formed dust at some point. For V339 Del, the hydrogen Balmer lines were initially flat-topped and later developed  a  more ``peaked" profile, which is opposite to the evolution of nova SMCN 2016-10a. In the latter case, the peak at $\sim$ + 1200 km s$^{-1}$ faded and the lines show jagged, flat-topped profiles. These jagged, flat-topped profiles are probably an indication of clumpiness in the ejecta, which might lead to a disordered magnetic field and therefore reduce the intrinsic polarization (see Section~\ref{spectropol}). They might also be associated with the complete envelope ejection at the end of the wind phase after the eruption \citep{Della_Valle_etal_2002}. The evolution of the lines also resemble that of nova V959 Mon (nova Mon 2012) in \citet{Shore_etal_2013}. Nova V959 Mon is a slow ONe nova and it was the first nova to be discovered in the $\gamma$-rays prior to optical detection due to its proximity to the Sun during eruption. The Balmer lines of Nova V959 Mon show a red-shifted single emission peak with jagged profiles that evolved into more flat-topped profiles with time (see figure 2 in \citealt{Shore_etal_2013}).

\begin{figure}
  \includegraphics[width=\columnwidth]{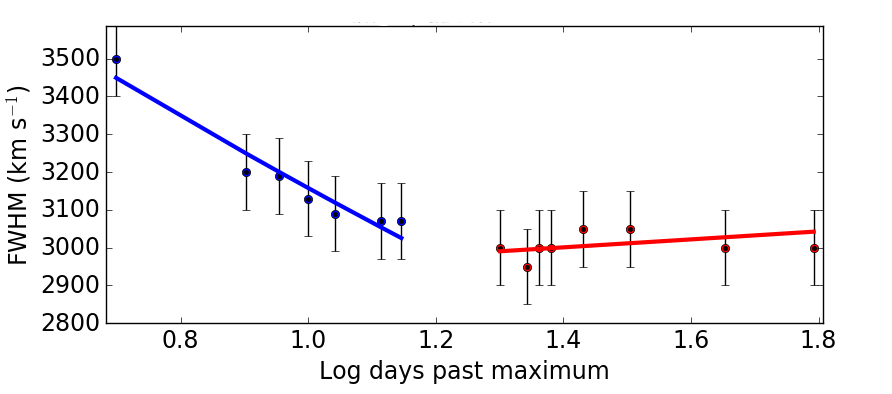}
\caption{The FWHM of H${\alpha}$ plotted against the $\log$ time in days since maximum light. The blue circles represent the FWHM before day 14 and the red circles represent the FWHM from day 20 onward. A power law was fitted to the evolution of the FWHM of H${\alpha}$ from day 5 to day 14 resulting in an exponent of $\sim$ -0.14.}
\label{Fig:FWHM}
\end{figure}
The FWHM of H${\alpha}$ and H${\beta}$ were derived by applying multiple component Gaussian fitting in the IRAF and python environments separately (see Figure~\ref{Fig:Balmerlines}). The results are shown in Table~\ref{tableFWHM}. The FWHM of both lines shows a decreasing trend and a systematic narrowing of the lines might be due to the deceleration of the ejecta after the eruption. The evolution of the FWHM is presented in Figure~\ref{Fig:FWHM}. This evolution follows a decreasing trend of a power law in time with an exponent n $\sim -$0.14 ($\Delta t^n$) where $\Delta t$ is equal to the time since maximum light.
Other novae have shown similar line narrowing after the eruption (see e.g. \citealt{Della_Valle_etal_2002,Hatzidimitriou_etal_2007,Shore_etal_2013,Darnley_etal_2016}).

\citet{Bode_etal_1985} carried out modelling for the 1985 eruption of RS Ophiuchi and they suggested a three-phase shock model, where the high-velocity ejecta interact with the low-velocity stellar wind from the companion, which is a red giant in the case of RS Ophiuchi. The interaction between the ejecta and the wind generates shocks that are responsible for the ejecta deceleration, and hence the line narrowing. 

Another explanation for this line narrowing was proposed by \citet{Shore_etal_1996}, who suggested that it is due to the distribution of the matter velocity at the ejection moment. The emissivity of the fastest moving gas decreases with time at a higher rate than the slower moving gas, due to the decrease of its density, leading to the line narrowing.

\begin{table}
\caption{The FWHM of H${\alpha}$ and H${\beta}$ at each observation. The values are in km s$^{-1}$.}
\begin{center}
\begin{tabular}{rrr}
\hline
\centering
$\Delta t = t-t_0$ & FWHM (H${\alpha}$)  & FWHM (H${\beta}$) \\ 
(days) &\multicolumn{2}{c}{ $\pm$ 100 (km s$^{-1}$)}  \\
\hline
5 & 3500 & 3450 \\
8 & 3200 & 3050 \\
9  & 3190 & 3050 \\
10 & 3130 & 3000 \\
11 & 3090 & 2850 \\
13 & 3070 & 2950 \\
14 & 3070 &  --\\
20 & 3000 & 3150 \\
22 & 2950 & 2900 \\
23 & 3000 & 2850 \\ 
24 & 3000 & 2900 \\ 
27 & 3050 & 3000 \\ 
32 & 3050 & 2950 \\ 
45 & 3000 & 2850 \\ 
62 & 3000 & 2800 \\
 \hline
\end{tabular}
\end{center}
\label{tableFWHM}
\end{table}

After a survey of novae in M31, \citet{Shafter_etal_2011} carried out spectroscopic and photometric analysis of 91 novae and derived a relation between the FWHM of H${\alpha}$ and the light-curve parameter $t_2$ showing a clear trend of faster novae exhibiting higher expansion velocities. Although, the study is conducted in a different galaxy than the SMC, which means different metallicity and Hubble type, we applied their relation to this nova: 
$$\log t_2 (d) = 6.84 \pm 0.1 - (1.68 \pm 0.02) \times \log [\mathrm{FWHM_{H{\alpha}}} (\mathrm{km\,s^{-1})}],$$
with a FWHM of $\sim$ 3500 $\pm$ 100\,km\,s$^{-1}$,  leading to $t_2 = 7.6 \pm 2$ days which is slightly larger than the values derived from our photometric data (see section~\ref{LC}).

\begin{figure}
\begin{center}
  \includegraphics[width=0.65\columnwidth]{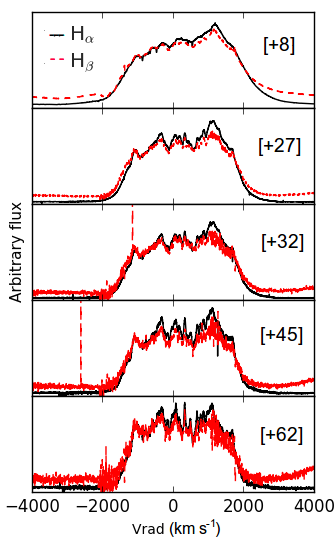}
\caption{The evolution of H$\alpha$ (solid black) and H$\beta$ (dashed red) line profiles. The numbers between brackets are days since $t_0$.}
\label{Fig:halphabeta}
\end{center}
\end{figure}

In order to check for the blending of the \feal{N}{II} 6548\,$\mathrm{\AA}$ and 6583\,$\mathrm{\AA}$ lines with H$\alpha$ in terms of an apparent extra emission component in radial velocity, we directly compared the evolution of H$\alpha$ and H$\beta$ (Figure~\ref{Fig:halphabeta}). Although the lines are very broad, no evidence of blending is present in any of the observations. If present, the \feal{N}{II} 6548\,$\mathrm{\AA}$ is at $\sim$ - 600\,km\,s$^{-1}$ to the blue side of H$\alpha$ and \feal{N}{II} 6583\,$\mathrm{\AA}$ is at $\sim$ + 1000\,km\,s$^{-1}$ to the red side of H$\alpha$. The line profile structure and evolution of H$\alpha$ and H$\beta$ is almost identical in the five high resolution observations. H$\alpha$ does not show any extra emission at these two velocities in comparison to H$\beta$. The peak at $\sim$ + 1200\,km\,s$^{-1}$ is present in both lines, as well in H$\delta$ and H$\gamma$ as discussed previously. 

\subsubsection{Oxygen lines}
\begin{figure}
  \includegraphics[width=\columnwidth]{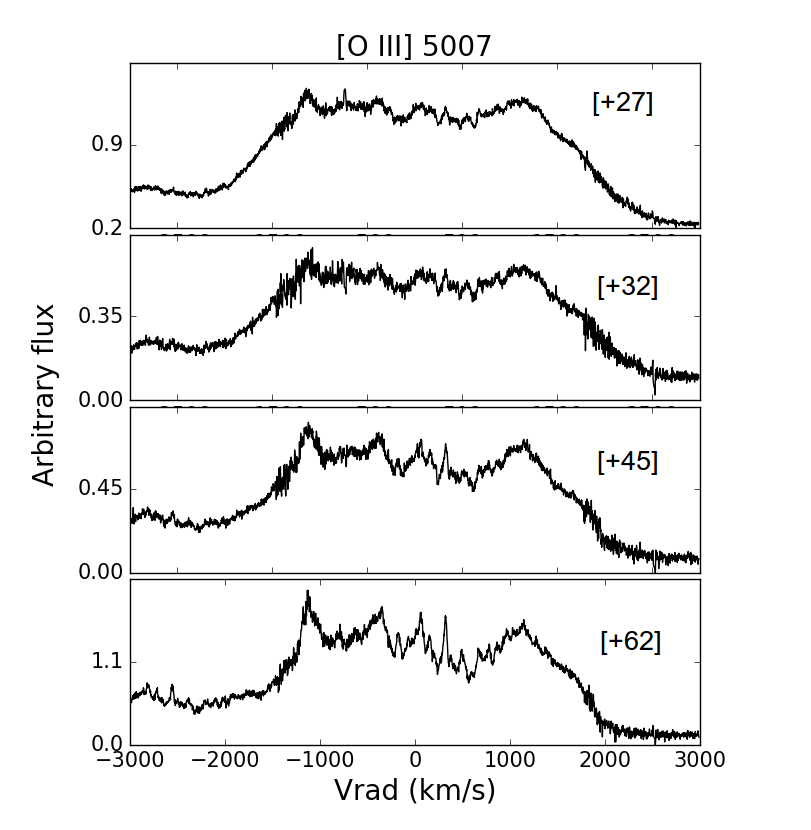}
\caption{The evolution of \feal{O}{III} 5007\,$\mathrm{\AA}$ line profiles. The rise to the blue side of the line is due to the \feal{O}{III} 4959\,$\mathrm{\AA}$ line. The numbers between brackets are days after $t_0$. Heliocentric correction is applied to the radial velocities.}
\label{Fig:[OIII]}
\end{figure}

The nebular \feal{O}{III} 4363\,$\mathrm{\AA}$, 4959\,$\mathrm{\AA}$, and 5007\,$\mathrm{\AA}$ lines were observed in all the spectra from day 20 onwards. The lines were absent in the first two spectra (day 5 and 8) since the nova was still close to maximum light and only entered the nebular stage in late October.  The \feal{O}{III} 4363\,$\mathrm{\AA}$ is blended with H$\gamma$. We show the evolution of the \feal{O}{III} 5007\,$\mathrm{\AA}$ line in Figure~\ref{Fig:[OIII]}.

\citet{Williams_2012} showed that the \eal{O}{I} 8446/7773 intensity ratio is a good indicator of the density of gas that forms the spectrum during early decline. \eal{Fe}{II} spectra are expected to have a high intensity 7773\,$\mathrm{\AA}$ line, which is thought to originate from high density gas that persists for months after the eruption. Hence, the \eal{O}{I} 8446/7773 ratio is low for the \eal{Fe}{II} spectra (sometimes lower than 2), and sometimes showing small changes over months time. For the He/N class, the lines are expected to originate from low density gas where the line at 7773\,$\mathrm{\AA}$ is sometimes not even detected. Hence, the ratio of the two oxygen lines should be high, followed by a fast decrease. We derived the flux ratio of \eal{O}{I} 8446/7773 for all our observations and the results are shown in Table~\ref{tableOI}. They show a high increasing ratio followed by a fast decrease, as expected for a He/N nova.
\begin{table}
\centering
\caption{The flux ratio of \eal{O}{I} 8446/7773 from the HRS and RSS observations. The flux is not corrected for reddening.}
\begin{tabular}{rr}
\hline
$\Delta t = t-t_0$ &  $F$(\eal{O}{I}8446/7773) \\ 
(days) & \\
\hline
5 & 2.3\\
8 & 3.6\\
23 & 5.5\\ 
27 & 5.4\\ 
32 & 4.5\\ 
45 & 2.3\\ 
62 & 1.2\\ 
\hline
\end{tabular}
\label{tableOI}
\end{table}

\section{Summary and conclusions}
\label{Conc}
Nova SMCN 2016-10a was discovered on 2016-10-14.19 UT by MASTER-OAFA auto-detection system \citep{ATel_9621}. Pre-discovery observations from MASTER \citep{ATel_9631} and a DSLR camera from Sao Jose dos Campos, Brazil \citep{ATel_9684}, allowed us to estimate $t_0$ as 2016-10-09.2 UT and the date of maximum optical light as 2016-10-09.8 UT.  Nova SMCN 2016-10a is the best studied nova in the SMC so far. The optical, NIR, X-ray, and UV data, led us to the following conclusions:
\begin{enumerate}[1.]
\item Nova SMCN 2016-10a is a very fast nova with $t_2\simeq$ 4.0\,$\pm$\,1.0\,d. This is an indication of an eruption that occurred on a high mass WD ($M_{\mathrm{WD}}$\,$\geq$\,1.25\,M$_{\odot}$).
\item The light-curve is consistent with a nova at the distance of the SMC, although its magnitudes at 15 and 17 days indicate it may be to the fore of the bulk of the SMC. 
\item At a distance of 61\,$\pm$\,10\,kpc, it is the brightest nova discovered in the cloud with $M_{V,\mathrm{max}}$ $\approx$ $-$10.5\,$\pm$\,0.5 and probably one of the brightest on record.
\item Based on the photometric data from the quiescence phase and the early decline phase, we suggest that the progenitor system is likely to contain a main sequence secondary, or a sub-giant secondary.
\item Based on the optical spectra, we classified nova SMCN 2016-10a as a He/N nova indicating a system with a high mass WD.
\item After 5 days from optical peak, we measured FWHM of\,$\sim$\,3500\,$\pm$\,100\,km\,s$^{-1}$ from the H${\alpha}$ and H${\beta}$ lines, indicating moderately high expansion velocities.
\item The light-curve during quiescence, the Balmer line profiles, as well as the negligible polarization, suggest the system might be at a low inclination and that there are clumps in the ejecta. Line modelling is definitely needed to give us a better insight into the structure and morphology of the system and the ejecta. 
\item The high temperature ($\sim$90\,eV) of the super-soft X-ray emission, together with the relatively rapid turn-on and -off times (28\,d and 180\,d, respectively), suggest a high mass WD ($\sim$1.25-1.3 M$_{\odot}$), in agreement with the results
from the optical data. We note that, at a distance of 61\,kpc, the bolometric luminosity is a factor of 1.76 higher than at 46\,kpc.
\item The plateaux in the UV light-curves extend at least from day $\sim$90 to 170,
approximately corresponding to the interval of bright, high temperature soft X-rays.
\item The UV line fluxes show different light-curves before day 20 suggesting optical thickness is the cause. The line fluxes have the same light-curves in the later data, suggesting the ejecta are effectively optically thin by then, and that is also when we see the rise of the X-ray flux. 
\end{enumerate}  

Once the system returns to quiescence, more observations are essential to constrain the type of the secondary star. If it becomes possible to measure the expansion parallax of this nova then the question of its brightness and location could be settled. Such measurements are not easy and often raise questions about the symmetry of the eruption. With the available spectroscopic data, line modelling is encouraged to follow-up this work, which can help to constrain the structure and morphology of the system.

\section*{Acknowledgments}
A part of this work is based on observations made with the Southern African Large
Telescope (SALT), under the program 2016-1-MLT-010 and 2016-2-LSP-001. EA, DB, PAW, SM, PW gratefully acknowledge the receipt of research grants from the National Research Foundation (NRF) of South Africa.  We are grateful to Steve Crawford, Itumeleng Monageng, Daniel Viljoen, and Brent Miszalski for assistance with the SALT observations\\
AK acknowledges the National Research Foundation of South Africa and the Russian Science Foundation (project no.14-50-00043).\\
P.M. is supported by the ''Diamond Grant'' No. DI2013/014743 funded by the Polish Ministry of Science and Higher Education. \\
The OGLE project has received funding from the National Science Center, Poland, grant MAESTRO 2014/14/A/ST9/00121 to A.U.\\
KLP, NPMK, APB and JPO acknowledge support from the UK Space Agency.\\
FMW acknowledges support from the US National Science Foundation, grant 1614113. Based in part on observations from the SMARTS Observatory, which is hosted by the Cerro Tololo Inter American Observatory, National Optical Astronomy Observatory, which is operated by the Association of Universities for Research in Astronomy (AURA) under a cooperative agreement with the National Science Foundation. We thank the SMARTS queue schedulers and observers for their efforts.\\
Based on observations obtained at the Southern Astrophysical Research (SOAR) telescope, which is a joint project of the Minist\'{e}rio da Ci\^{e}ncia, Tecnologia, e Inova\c{c}\~{a}o (MCTI) da Rep\'{u}blica Federativa do Brasil, the U.S. National Optical Astronomy Observatory (NOAO), the University of North Carolina at Chapel Hill (UNC), and Michigan State University (MSU).\\
VARMR acknowledges financial support from FCT in the form of an exploratory project of reference IF/00498/2015, from CIDMA strategic project UID/MAT/04106/2013 and supported by Enabling Green E-science for the Square Kilometer Array Research Infrastructure (ENGAGE SKA), POCI-01-0145-FEDER-022217, funded by Programa Operacional Competitividade e Internacionaliza\c{c}\~{a}o (COMPETE 2020) and FCT, Portugal.\\
We thank Ulisse Munari and an anonymous referee for useful comments.\\
\bibliography{biblio}

\appendix
\section{Observations log}
\label{app}
In this Appendix we list all the observations log. The OGLE, SMARTS, and {\em Swift} UVOT photometry can be found on the electronic version. The OGLE time series photometry are also available from the OGLE Internet Archive\footnote{http://ogle.astrouw.edu.pl/ogle4/cvom/smcn-2016-10a.html}. The SMARTS time series photometry is available in the SMARTS atlas, at \footnote{http://www.astro.sunysb.edu/fwalter/SMARTS/NovaAtlas/nsmc2016/nsmc2016.html}. 
In Table~\ref{line_det} we list the line IDs along with the FWHM, Equivalent Widths (EWs), and fluxes for those emission lines for which an estimate was possible.
\clearpage

\begin{table}
\centering
\caption{A sample of OGLE $V$-band photometry. The time series photometry is available from the OGLE Internet Archive and on the electronic version.}
\begin{tabular}{rrrr}
\hline
HJD  & $t-t_0$  & $V$ & $\Delta V$\\ 
  & (days) & \multicolumn{2}{c}{(mag)}\\
\hline
2455399.88 &  $-$2270.81 & 20.60 & 0.03\\
2455434.82 &  $-$2235.87 & 20.87 & 0.12\\
2455446.70 &  $-$2223.99 & 20.47 & 0.02\\
2455456.81 &  $-$2213.88 & 20.58 & 0.03\\
2455486.62 &  $-$2184.07 & 20.40 & 0.06\\
2455499.63 &  $-$2171.07 & 20.98 & 0.04\\
2455503.62 &  $-$2167.08 & 20.72 & 0.04\\
2455516.60 &  $-$2154.10 & 20.51 & 0.06\\
2455536.58 &  $-$2134.11& 20.48 & 0.03\\
 \hline
\end{tabular}
\label{table:OGLEV}
\end{table}

\begin{table}
\centering
\caption{A sample of OGLE $I$-band photometry. The time series photometry is available from the OGLE Internet Archive and on the electronic version.}
\begin{tabular}{rrrr}
\hline
 HJD  & $t-t_0$ & $I$ & $\Delta I$\\ 
 & (days)&  \multicolumn{2}{c}{(mag)}\\
\hline
2455346.93 & $-$2323.77 &20.14 & 0.20\\
2455347.91 & $-$2322.79 &20.42 & 0.23\\
2455358.90 & $-$2311.80 &20.06 & 0.13\\
2455364.91 & $-$2305.79 &20.35 & 0.13\\
2455376.94 & $-$2293.76 &20.09 & 0.13\\
2455378.94 & $-$2291.76 &20.40 & 0.17\\
2455380.91 & $-$2289.79 &20.35 & 0.20\\
2455381.92 & $-$2288.79 &20.44 & 0.21\\
2455384.89 & $-$2285.81 &20.49 & 0.20\\
 \hline
\end{tabular}
\label{table:OGLEI}
\end{table}

\begin{table}
\centering
\caption{The FLOYDS spectroscopic observations log.}
\begin{tabular}{rrr}
\hline
 HJD & $t - t_0$ &Exposure time  \\ 
 &  (days) &  (s) \\
\hline
2457676.19 & 5.4  & 3$\times$60\\
2457679.10 & 8.3  &3$\times$60\\
2457680.04 & 9.3 & 3$\times$60\\
2457681.20 & 10.5 & 6$\times$60\\
2457682.00 & 11.3 & 3$\times$120\\
2457684.12 & 13.4  & 2$\times$120\\
2457685.13 & 14.4  & 3$\times$120\\
2457691.11 & 20.4 & 3$\times$300\\
2457693.15 & 22.4  & 3$\times$300\\
2457695.07 & 24.3 & 3$\times$450\\
2457703.20 & 32.5 & 3$\times$450\\
\end{tabular}
\label{tablefts}
\end{table}

\begin{table}
\centering
\caption{A sample of SMARTS \textit{BVRIJHK} photometry. The time series photometry is available on the electronic version and on the SMARTS atlas.}
\begin{tabular}{rrrrr}
\hline
HJD & $(t - t_0)$   & Band & Magnitude & Instrument\\ 
&(days) & & & uncertainty\\
\hline
2457682.64 &11.93 & $I$&   10.84 & 0.002\\
2457682.64 &11.93 & $H$&  10.21&  0.030\\ 
2457682.64 &11.93 & $B$&  11.73&  0.001\\ 
2457682.64 &11.93 & $J$&   9.83 & 0.037 \\
2457682.64 &11.93 & $K$&   9.95 & 0.033 \\
2457682.64 &11.93 & $V$&  12.14&  0.002 \\
2457682.64 &11.93 & $R$&  10.64&  0.001 \\
2457683.73 &13.02 & $I $&  10.90 & 0.002 \\
2457683.73 &13.02 & $J$&   9.95 & 0.038 \\
2457683.73 &13.02 & $H$&  10.45&  0.034 \\
2457683.73 &13.02 & $B$&  11.57&  0.001 \\
2457683.73 &13.02 & $K$&  10.05&  0.039 \\
2457683.73 &13.02 & $V$&  12.07&  0.002 \\
2457683.73 &13.02 & $R$&  10.67&  0.001 \\
2457684.73 &14.02 & $I $& 11.10 & 0.002 \\
2457684.73 &14.02 & $J$&  10.74&  0.043 \\
2457684.73 &14.02 & $H$&  10.653&  0.036 \\
2457684.73 &14.02 & $B$&  11.73&  0.001 \\
2457684.73 &14.02 & $K$&  10.02&  0.04\\
\end{tabular}
\label{table:SMARTS}
\end{table}

\begin{figure*}
\begin{center}
  \includegraphics[width=\textwidth]{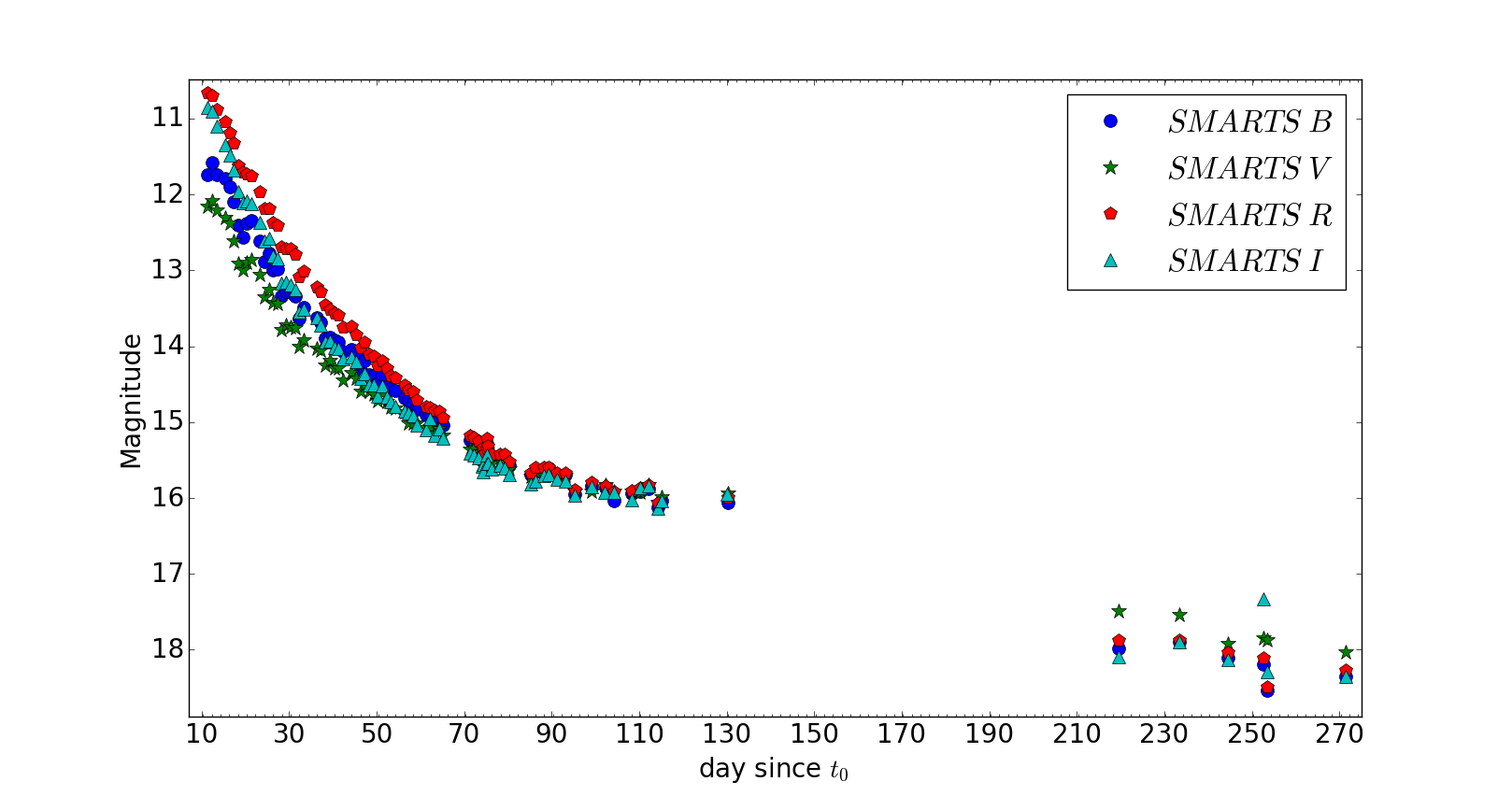}
\caption{The full photometric $BVRI$ data from SMARTS as a function of days since eruption, colour and symbol coded as indicated in the legend. A colour version of this plot is present in the online journal.}
\label{Fig:BVRI_LC}
\end{center}
\end{figure*}

\begin{table}
\centering
\caption{A sample of $Swift$ UVOT $uvw2$ photometry. The time series photometry is available on the electronic version.}
\begin{tabular}{rrrrr}
\hline
 HJD  & $t-t_0$ & exposure time & $uvw2$ & $\Delta uvw2$\\ 
 & (days)& (s) &  \multicolumn{2}{c}{(mag)}\\
\hline
2457682.85& 12.15 &4.8 & 9.85& 0.14\\
2457682.86& 12.16&4.8 & 9.68& 0.21\\
2457682.92& 12.22&4.8 & 9.64& 0.14\\
2457682.93& 12.23&16.4 & 9.76& 0.18\\
2457685.18& 14.48&4.8 & 10.65& 0.43\\
2457685.18& 14.48& 15 & 10.12& 0.15\\
2457685.24& 14.54& 15 & 10.15& 0.15\\
2457685.25& 14.55& 10.7 & 9.93& 0.215\\
2457685.57& 14.87& 4.8& 10.07& 0.15\\
2457685.58& 14.88&  17 & 10.06& 0.19\\
 \hline
\end{tabular}
\label{table:uvw2}
\end{table}

\begin{table}
\centering
\caption{A sample of $Swift$ UVOT $uvm2$ photometry. The time series photometry is available on the electronic version.}
\begin{tabular}{rrrrr}
\hline
 HJD  & $t-t_0$ & exposure time & $uvm2$ & $\Delta uvm2$\\ 
 &  (days) & (s)&  \multicolumn{2}{c}{(mag)}\\
\hline
2457682.85 & 12.15 & 4.8 & 9.61 & 0.27\\
2457682.85 & 12.15 & 50 & 9.91 & 0.15\\
2457682.92 & 12.22 & 4.8 & 9.76 & 0.30\\
2457685.18 & 14.48 & 4.8 & 10.12 & 0.15\\
2457685.25 & 14.55 & 4.8 & 10.26 & 0.16\\
2457685.57 & 14.87 & 4.8 & 10.32 & 0.03\\
2457703.33 & 32.63 & 4.8 & 12.03 & 0.05\\
2457703.33 & 32.63 & 46 & 12.07 & 0.03\\
 \hline
\end{tabular}
\label{table:uvm2}
\end{table}

\begin{table*}
\begin{minipage}{150mm}
\caption{Summed $Swift$ UV grism exposures }
\label{grismobs}
\begin{tabular}{@{}rrrl}
\hline
HJD   &  HJD  &   $ <t-t_0> $  &  quality notes$^1$\\
(start) & (end) & (days) & \\
\hline 
2457676.8036 & 2457678.7299 &   7.07 & 2ndO:2870-2950,3308-3390,3890-3980,4400-4500,5080-5150\\
2457682.8616 & 2457685.2581 &  13.36 & Good 1700-4500,but rise above 4500-5060\\
2457685.5775 & 2457688.7052 &  16.44 & Good 1700-5060,5060-6000 noisy and strange drop in flux\\
2457699.5597 & 2457705.4783 &  31.82 & ZOc:1831-1885;2ndO:2824-2945,3235-3375,3843-3947,4375-4530,5050-5170\\
2457710.9143 & 2457714.0447 &  41.78 & Good 1700-4800, 4800-5100 shows rising continuum to red\\
2457715.6864 & 2457719.7391 &  47.01 & Good 1700-5200, then noisy\\
2457723.6584 & 2457734.8146 &  58.54 & Good 1700-5100 then noisy\\
2457762.1341 & 2457776.4872 &  98.61 & Continuum level good to 3840 \\
2457788.4264 & 2457828.3827 & 137.70 & No contamination, bad S/N, very noisy for $>$4100\\
\hline
\end{tabular}
$^1$ 2ndO: second order affects these ranges; ZOc: zeroth order contamination affects these ranges.\\
\label{uvgrismtable}
\end{minipage}  
\end{table*}

\begin{table*}
\centering
\caption{The SALT spectroscopy and spectropolarimetry observations log.}
\begin{tabular}{rrrrr}
\hline
 Type & HJD & $ t - t_0$ & Exposure time (range 1) & Exposure time (range 2)  \\ 
& & (days) &  (s) & (s)\\
\hline
RSS & 2457679.4 & 8 & 4$\times$150 & 3$\times$150 + 4$\times$30\\
RSS & 2457694.3 & 23 & 2$\times$150 + 2$\times$30 & 2$\times$100 + 2$\times$30\\
HRS & 2457679.3 & 8 & 3$\times$1200 & 2$\times$1200\\
HRS & 2457698.3 & 27 & 3$\times$1200 & 2$\times$1200\\
HRS & 2457703.3 & 32 & 3$\times$1200 & 2$\times$1200\\
HRS & 2457716.3 & 45 & 3$\times$1200 & 2$\times$1200\\
HRS & 2457733.2 & 62 & 3$\times$1200 & 2$\times$1200\\
Spectropol & 2457686.3 & 15 & 8$\times$120 & 8$\times$120 + 4$\times$60\\
Spectropol & 2457714.3 & 43 & 8$\times$120 & 8$\times$120 + 4$\times$60\\
\end{tabular}
\label{table:SALT}
\end{table*}

\begin{table}
\centering
\caption{Line IDs, EWs, FWHMs, and fluxes of some emission lines in the nova SMCN 2016-10a spectra.}
\begin{tabular}{rrrrrr}
\hline
Line & $\lambda_0$  & EW ($\lambda$) & FWHM  & Flux & ($t-t_0$)\\ 
 & \multicolumn{2}{c}{($\mathrm{\AA}$)} & (km\,s$^{-1}$) & erg cm$^{-2}$s$^{-1}$\\
\hline
\feal{Ne}{V} & 3426 & - & - & - & 32 \\
H$\eta$ & 3835 & - & - & - & 8\\
H$\zeta$ & 3889 & $-$11$\pm$2 &2250$\pm$100 & - & 8\\
H$\epsilon$ & 3970 & $-$15$\pm$10 &2300$\pm$100 & 1.54$\times$10$^{-12}$ & 8\\
H$\delta$ & 4102 & $-$85$\pm$10 &3300$\pm$100 &2.24$\times$10$^{-12}$ & 8\\
H$\gamma$ & 4341 & $-$100$\pm$10 &3450$\pm$100 &2.48$\times$10$^{-12}$ & 8\\
\eal{He}{I} & 4388 & - & - & - & 62\\
\eal{N}{III} & 4638 & - & - & - & 8\\
\eal{He}{II} & 4686 & - & - & - & 8\\
H$\beta$ & 4861 & $-$300$\pm$20 &3500$\pm$100 & 4.73$\times$10$^{-12}$ & 8\\
\feal{O}{III} & 4959 & - & - & - & 20\\
\eal{N}{II} & 5001 & - & - & - & 5\\
\feal{O}{III} & 5007 & $-$290$\pm$20 & 3000$\pm$100 & 1.59$\times$10$^{-13}$ & 20\\
\eal{He}{I} & 5016 & - & - & - & 5\\
\eal{Fe}{II} & 5018 & - & - & - & 5\\
\eal{Fe}{II} & 5169 & - & - & -  & 5\\
\eal{N}{II} & 5679 & - & - & - & 8\\
\feal{N}{II} & 5755 & - & - & - & 8\\
\eal{He}{I} & 5876 & - & - & - & 8\\
\eal{Na}{I} D & 5889 & - & - & - & 8\\
\eal{N}{II} & 5938 & - & - & - & 8\\
H$\alpha$ & 6563 & $-$1500$\pm$100 &3500$\pm$100 & 1.82$\times$10$^{-11}$ & 8\\
\eal{N}{I} & 7452 & - & - & - & 8\\
\eal{O}{IV} & 7713 & - & - & - & 62\\
\feal{C}{IV} & 7726 & - & - & - & 62\\
\eal{O}{I} & 7773 & $-$135$\pm$10 & 2400$\pm$100 & 1.83$\times$10$^{-12}$ & 8\\
\eal{Mg}{II} & 7896 & - & - & - & 8\\
\eal{N}{I} & 8212 & - & - & - & 8\\
\eal{Mg}{II} & 8232 & - & - & - & 8\\
\eal{O}{I} & 8446 & $-$750$\pm$50 & 3000$\pm$100 & 4.00$\times$10$^{-12}$ & 8\\
\eal{N}{I} & 8617 & - & - & - & 8\\
\eal{N}{I} & 8692 & - & - & - & 8\\
\hline
\end{tabular}
\label{line_det}
\end{table}

\label{lastpage}
\end{document}